\documentclass[12pt]{article}
\usepackage{epsfig,citesort,enumerate,amsmath,fullpage,color,latexsym}
\def\myendproof{{\ \vbox{\hrule\hbox{%
   \vrule height1.3ex\hskip0.8ex\vrule}\hrule }}\par}
\newtheorem{theorem}{Theorem}[section]
\newtheorem{lemma}[theorem]{Lemma}

\newenvironment{proof}{{\it Proof. }}{\myendproof}
\newcommand{\select}[1]{\text{select}({#1})}
\newcommand{\rank}[1]{\text{rank}({#1})}
\newcommand{\match}[1]{\text{match}({#1})}
\newcommand{\enclose}[1]{\text{enclose}({#1})}
\newcommand{\enclosek}[2]{\text{enclose}_{#1}({#2})}
\newcommand{\level}[1]{\ell({#1})}
\newcommand{\son}[1]{\text{wrapped}({#1})}
\newcommand{\rev}[1]{\text{reverse}({#1})}
\newcommand{\floor}[1]{{\left\lfloor{#1}\right\rfloor}}
\newcommand{\ceiling}[1]{{\left\lceil{#1}\right\rceil}}
\newcommand{\leftp}[1]{{\mbox{\tt (}_{#1}}}
\newcommand{\rightp}[1]{{\mbox{\tt )}_{#1}}}
\newcommand{\leftb}[1]{{\mbox{\tt [}_{#1}}}
\newcommand{\rightb}[1]{{\mbox{\tt ]}_{#1}}}
\newcommand{\aux}{\chi}
\newcommand{\algo}[1]{\mbox{\sf{#1}}}
\newcommand{\setof}[1]{\{{#1}\}}
\newcommand{\set}[2]{\{{#1}\mid{#2}\}}
\newcommand{\Xomit}[1]{}
\newcommand{\next}[1]{\mbox{next}({#1})}
\newcommand{\prev}[1]{\mbox{prev}({#1})}
\newcommand{\leaf}[1]{\text{leaf}(T'_{#1})}
\newcommand{\kfrac}[0]{\left( 5+\frac{1}{k}\right)}
\newtheorem{fact}[theorem]{Fact}
\newcommand{\block}{\algo{block}{}}
\newcommand{\draw}{\algo{draw}{}}

\title{Orderly Spanning Trees with Applications\thanks{
A preliminary version appeared in {\em Proceedings of the $12$-th
Annual ACM-SIAM Symposium on Discrete Algorithms}, Washington,
D.C., USA, January 7--9, 2001, pp.~506--515. This research is
supported in part by NSC grants NSC 89-2213-E-001-034 and NSC
89-2218-E-001-014.} }
\author{%
Yi-Ting Chiang
\and
Ching-Chi Lin
\and
Hsueh-I Lu\thanks{Corresponding author.
Address: 128 Academia Road, Section 2, Taipei 115, Taiwan. Affiliation: Institute of Information Science, Academia Sinica. Phone: +886-2-2788-3799 ext 1731. eFax: +1-708-570-9151. 
URL: www.iis.sinica.edu.tw/\~{ }hil/.
Email: hil@iis.sinica.edu.tw.}
}

\date{Academia Sinica, Taiwan}
\begin{document}

\maketitle

\begin{abstract}
We introduce and study the {\em orderly spanning trees} of plane
graphs.  This algorithmic tool generalizes {\em canonical orderings},
which exist only for triconnected plane graphs.  Although not every
plane graph admits an orderly spanning tree, we provide an algorithm
to compute an {\em orderly pair} for any connected planar graph $G$,
consisting of a plane graph $H$ of $G$, and an orderly spanning tree
of $H$.  We also present several applications of orderly spanning
trees: (1) a new constructive proof for Schnyder's Realizer Theorem,
(2) the first area-optimal 2-visibility drawing of $G$, and (3) the
best known encodings of $G$ with $O(1)$-time query support. All
algorithms in this paper run in linear time.
\end{abstract}

\section{Introduction}
The {\em canonical orderings} of triconnected plane
graphs~\cite{DeFPP90,Kant96,KantH98,HarelS98} are crucial in several
graph-drawing and graph-encoding
algorithms~\cite{FoessmeierKK97,HKL,ChuangGHKL98,He99,Chrobak95,ChrobakN98}.
This paper introduces an algorithmic tool {\em orderly spanning tree},
which generalizes the concept of canonical ordering for plane graphs
unrequired to be triconnected. Although not every connected plane
graph admits an orderly spanning tree, we provide a linear-time
algorithm to compute an {\em orderly pair} for any connected planar
graph $G$, consisting of a plane graph $H$ of $G$, and an orderly
spanning tree of $H$.

\subsection{Application 1}
For the first application of orderly spanning trees, we present a new
linear-time algorithm to compute a {\em realizer} for any plane
triangulation (i.e., simple triangulated plane graph with at least
three nodes). Schnyder~\cite{Schnyder89} gave the first known
linear-time algorithm that computes a realizer for any plane
triangulation, and thus, settling the open question on the
dimension~\cite{trotter92,Dushnik41} of planar graphs. This celebrated
result also yields the best known straight-line drawing of planar
graphs on the grid~\cite{Schnyder90}. The original proof of Schnyder's
Realizer Theorem is complicated.  
Our proof, based upon the existence of orderly spanning tree for any
simple plane triangulation, is relatively simple.

\subsection{Application 2}
\begin{figure}
\begin{center}
\input{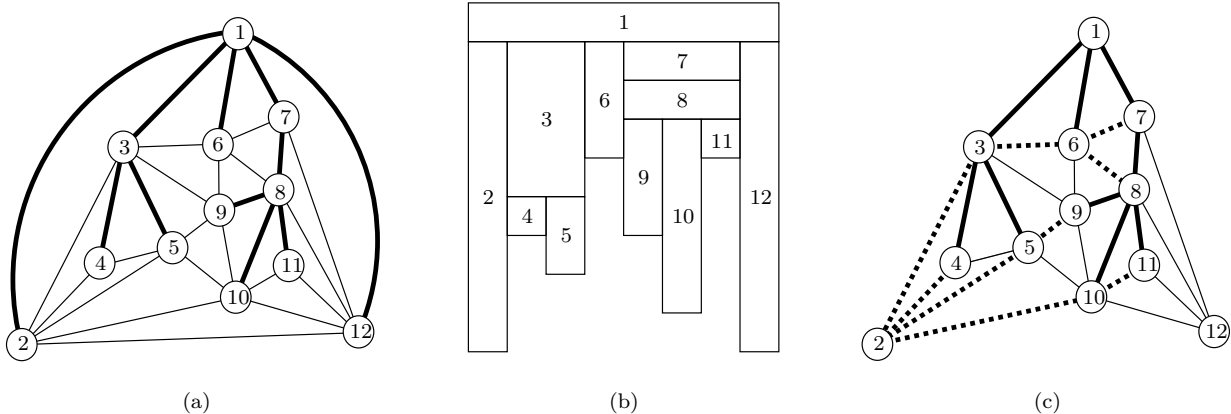}
\end{center}
\caption{(a) A plane graph $H$ with an orderly spanning tree of $H$
rooted at node 1 represented by the thick edges.  (b) A 2-visibility
drawing of $H$. (c) A realizer $(T_1,T_2,T_{12})$ of $H$, where $T_1$
(respectively, $T_2$ and $T_{12}$) consists of the thick
(respectively, dashed and thin) edges.}
\label{fig:vibi-real}
\end{figure}

For the second application of orderly spanning trees, we give an
$O(n)$-time algorithm that produces a 2-visibility drawing for any
$n$-node simple plane graph $H$, with $n\geq 3$, whose area is at most
$(n-1) \times \floor{\frac{2n+1}{3}}$.  Let $v_1,v_2,\ldots,v_n$ be
the nodes of $H$.  A {\em 2-visibility drawing}~\cite{FoessmeierKK97}
of $H$ consists of $n$ non-overlapping rectangles $b_1,b_2,\ldots,b_n$
such that if $v_i$ and $v_j$ are adjacent in $H$, then $b_i$ and $b_j$
are visible to each other either horizontally or
vertically.\footnote{A closely related {\em rectangle-visibility
drawing}~\cite{DeanH98,DeanH97,HutchinsonSV99,BoseHS96} of $H$
requires that $v_i$ and $v_j$ are adjacent in $H$ {\em if and only if}
$b_i$ and $b_j$ are visible to each other.} For example, the picture
in Figure~\ref{fig:vibi-real}(b) is a 2-visibility drawing of the
plane graph in Figure~\ref{fig:vibi-real}(a). F\"{o}\ss{}meier, Kant,
and Kaufmann~\cite{FoessmeierKK97} gave an $O(n)$-time algorithm to
compute an $x\times y$ 2-visibility drawing for $H$ with $x+y\leq 2n$,
and conjectured that it is ``not trivial'' to improve their upper
bound.  Moreover, they showed an $n$-node plane triangulation whose $x
\times y$ 2-visibility drawing requires $x+y\geq
n-1+\floor{\frac{2n+1}{3}}$ and
$\min\setof{x,y}\geq\floor{\frac{2n+1}{3}}$.\footnote{The lower bounds
stated in~\cite{FoessmeierKK97} are $x+y\geq\frac{5n}{3}$ and
$\min\setof{x,y}\geq\frac{2n}{3}$. Based on the given sketch of proof,
however, it is not hard to see that their lower bound should be
corrected as $x+y\geq n-1+\floor{\frac{2n+1}{3}}$ and
$\min\setof{x,y}\geq\floor{\frac{2n+1}{3}}$.}  According to their
lower bounds, the 2-visibility drawing produced by our algorithm is
worst-case optimal.

In order to take advantage of the wonderful properties of canonical
orderings, many drawing algorithms work on triangulated versions of
input plane graphs. As pointed out in~\cite{HarelS98}, the initial
triangulation tends to ruin the original plane graph's structure.  Our
orderly-pair algorithm appears as a promising tool for drawing graphs
neatly and compactly, without first triangulating the given plane
graphs.  The concept of orderly pair is more general than that of
canonical ordering, since all known canonical orderings are defined
for plane graphs. The technique of orderly pairs is potentially more
powerful, since it exploits the flexibility of planar graphs whose
planar embeddings are not predetermined.

\subsection{Application 3}
For the third application of orderly spanning trees, we investigate
the problem of encoding a graph $G$ into a binary string $S$ with the
requirement that $S$ can be decoded to reconstruct $G$. This problem
has been extensively studied with three objectives: (1) minimizing the
length of $S$, (2) minimizing the time required to compute and decode
$S$, and (3) supporting queries efficiently. As these objectives are
often conflicting, a number of coding schemes with different
trade-offs have been proposed in the literature. The widely useful
adjacency-list encoding of an $n$-node $m$-edge graph $G$ requires
$2m\lceil\log_2 n\rceil$ bits.  Using the encoding schemes of Breuer
and Folkman~\cite{Breuer66,BreuerF67} developed during the 60's, the
adjacency of any two nodes can be determined by the Hamming distance
of their labels.  Talamo and Vocca~\cite{Talamo-Vocca-Compact-1998}
gave an encoding, obtainable in $O(n^3)$ time, that assigns an
$O(d\log^3n)$-bit label to each degree-$d$ node. Without accounting
for the time required to read the labels, the adjacency of two nodes
can be determined from their encoding in $O(1)$ time. For certain
graph families, Kannan, Naor, and Rudich~\cite{KNR92} provided schemes
encoding each node with $O(\log n)$ bits, and supporting the $O(\log
n)$-time testing of adjacency between any two nodes. Instead of using
Schnyder's Realizer Theorem, Grossi and Lodi~\cite{GrossiL98} improved
the results in~\cite{KNR92} for planar graphs by inventing an $O(n\log
n)$-time algorithm to decompose any planar graph into three
edge-disjoint forests.\footnote{The results of
Schnyder~\cite{Schnyder89,Schnyder90} immediately imply a linear-time
algorithm for decomposing any planar graph into three edge-disjoint
forests.}  Cohen, Di~Battista, Kanevsky, and
Tamassia~\cite{CohenDKT93} provided an $O(n^4m4^k/k^2)$-time and
linear-space encoding of a $k$-connected $G$, supporting $O(1)$-time
query on whether any two nodes are connected by $k+1$ node-disjoint
paths.  Jacobson~\cite{Jacobson89} gave an $\Theta(n)$-bit encoding
for a connected and simple planar $G$ to support traversal in
$\Theta(\log{n})$ time per node visited.

Under the model of unit-cost
RAM~\cite{BrodnikM00,Handbook,Clark96,FredmanW94,Thorup97,Thorup2000},
where operations such as read, write, and add on $O(\log n)$
consecutive bits take $O(1)$ time, an encoding $S$ of $G$ is {\em
weakly convenient}~\cite{ChuangGHKL98} if it takes (i) $O(m+n)$ time
to encode $G$ and decode $S$, (ii) $O(1)$ time to determine from $S$
the adjacency of any two nodes in $G$, and (iii) $O(d)$ time to
determine from $S$ the neighbors of a degree-$d$ node in $G$. If the
degree of a node can be determined from a weakly convenient $S$ in
$O(1)$ time, then $S$ is {\em convenient}~\cite{ChuangGHKL98}.  For a
planar $G$ having multiple edges but no self-loops, Munro and
Raman~\cite{MR97} gave the first nontrivial convenient encoding of $G$
with $2m+8n+o(m+n)$ bits. Their result is based on the four-page
decomposition of planar graphs~\cite{Yannakakis89} and auxiliary
strings, encoding an involved three-level data structure for any
string of parentheses.  For a planar $G$ that has (respectively, has
no) multiple edges, Chuang, Garg, He, Kao, and Lu~\cite{ChuangGHKL98}
improved the bit count to $2m+{\kfrac}n+o(m+n)$ (respectively,
$\frac{5}{3}m+{\kfrac}n+o(n)$) for any positive constant $k$. They
also provided a weakly convenient encoding of
$2m+\frac{14}{3}n+o(m+n)$ (respectively, $\frac{4}{3}m+5n+o(n)$) bits
for a planar $G$ that has (respectively, has no) multiple edges.
Based on our orderly-pair algorithm, in this paper we present the best
known convenient encodings for a planar $G$: If $G$ may (respectively,
does not) contain multiple edges, then the bit count of our encoding
is $2m+3n+o(m+n)$ (respectively, $2m+2n+o(n)$), which is even less
than that of the weakly convenient encodings of Chuang {\em et
al.}~\cite{ChuangGHKL98}. The bit counts are very close to Tutte's
information-theoretical lower bound of roughly $3.58m$ bits for
encoding connected plane graphs without any query
support~\cite{Tutte63b}.  The bit count of our encoding for a planar
$G$ without multiple edges matches that of the best known convenient
encoding for an outerplanar graph~\cite{MR97}.  Besides relying on the
orderly-pair algorithm, our results are also based on an improved
auxiliary string for a folklore
encoding~\cite{MR97,HKL99,ChuangGHKL98} of a rooted tree $T$. With the
auxiliary strings of Munro and Raman~\cite{MR97}, computing the degree
of a degree-$d$ node in $T$ requires $\Theta(d)$ time.  In this paper,
we present a nontrivial auxiliary string, in Lemma~\ref{lemma:son}, to
support the degree query in $O(1)$ time.

If one only needs to reconstruct $G$ with no query support, the code
length can be substantially shortened. For this case, Tur\'{a}n
\cite{turan84} used $4m$ bits for a planar $G$ that may have
self-loops; this bound was improved by Keeler and Westbrook
\cite{KW:encodings} to $3.58m$ bits.  They also provided coding
schemes for several important families of planar graphs. In
particular, they used $1.53m$ bits for a triangulated simple $G$, and
$3m$ bits for a connected $G$ free of self-loops and degree-one
nodes. For a simple triangulated (respectively, triconnected) $G$, He,
Kao, and Lu~\cite{HKL} improved the bit count to $\frac{4}{3}m+O(1)$
(respectively, $\frac{3}{2}(\log_2
3)m+O(1)$). Rossignac~\cite{Rossignac99} independently showed how to
encode a triangulated $G$ in $\frac{4}{3}m+O(1)$ bits. Although all
these encodings can be encoded and decoded in linear time, none of
them is known to be information-theoretically optimal. For example,
the information-theoretic tight bound for plane triangulations, given
by Tutte~\cite{Tutte62}, is roughly $1.08m$.  Recently, He, Kao, and
Lu~\cite{HKL99,HKL2000} proposed an $O(n \log n)$-time framework for
encoding a graph in information-theoretically optimal number of
bits. This framework is applicable to various classes of planar
graphs. Lu~\cite{Lu02:SODA} improved the framework to run in $O(n)$,
and showed that its applicable to all graphs with genus $o(n\log^2)$
as long as their genus embeddings are given.
For dense graphs and complement
graphs, Kao and Teng~\cite{kaot94.isaac} devised two compressed
representations from adjacency lists to speed up basic graph
techniques.  Papadimitriou and Yannakakis~\cite{PH86.encode} and
Galperin and Wigderson~\cite{GW83} investigated complexity issues
arising from encoding a graph by a small circuit that computes its
adjacency matrix. For labeled planar graphs, Itai and
Rodeh~\cite{IR82} gave an encoding of $\frac{3}{2} n \log n + O(n)$
bits. For unlabeled general graphs, Naor~\cite{naor90} gave an
encoding of ${\frac{1}{2}}n^2-n\log{n}+ O(n)$ bits.  For encodings of
sparse graphs that need support for efficient updates,
see~\cite{Brodal99e,MunroRS01}. For parallel encoding algorithms for
sparse graphs, see~\cite{ArikatiMZ97}. A book in preparation by
Spinrad~\cite{Spinrad-book} surveys implicit representations for
various graph classes.

\subsection{Recent applications}
Besides the applications presented in the present paper, our
orderly-pair algorithm also yields the following recent results: (a)
Improved compact distributed routing tables for any $n$-node
distributed planar network~\cite{Lu02}, improving the best previously
known design of Gavoille and Hanusse~\cite{GavoilleICALP99} by
reducing the worst-case table size count from $8n+o(n)$ bits to
$7.181n+o(n)$ bits, without increasing the time complexity of
preprocessing and query.  (b) A linear-time algorithm for constructing
compact floor-plans for plane triangulations~\cite{LiaoLY01}, which is
not only much simpler than the previous methods in the
literature~\cite{YeapS93,He99}, but also provides the first known
nontrivial upper bound on the floor-plan's area.  (c) Compact Podevs
drawings for plane graphs and an alternative proof for the sufficient
and necessary condition for a planar graph to admit a rectangular
dual~\cite{ChenLLY02}.

\subsection{Organization of the paper}
The rest of the paper is organized as follows.
Section~\ref{section:algorithm} gives the linear-time algorithm for
computing an orderly pair of any given planar graph, whose
applications are given in
Sections~\ref{section:realizer}--\ref{section:data-structure}.
Section~\ref{section:realizer} gives the linear-time algorithm for
computing a realizer of any given plane
triangulation. Section~\ref{section:drawing} shows the linear-time
algorithm for obtaining an area-optimal 2-visibility drawing of any
given plane graph.  Section~\ref{section:data-structure} presents the
best known convenient encodings for planar graphs.

\section{Orderly spanning trees for plane graphs}
\label{section:algorithm}
\subsection{Basics}
A graph is {\em simple} if it contains no multiple edges.  Unless
stated otherwise, all graphs in
Sections~\ref{section:algorithm}--\ref{section:drawing} are simple. A
{\em plane graph} of a planar graph $G$ is the graph $G$ equipped with
a fixed planar embedding of $G$.  Let $H$ be a plane graph.  The {\em
contour} of $H$ is the boundary of the external face of $H$. The nodes
and edges on the contour of $H$ are {\em external} in $H$; and the
other nodes and edges are {\em internal} in $H$.

\begin{figure}
\begin{center}
\input{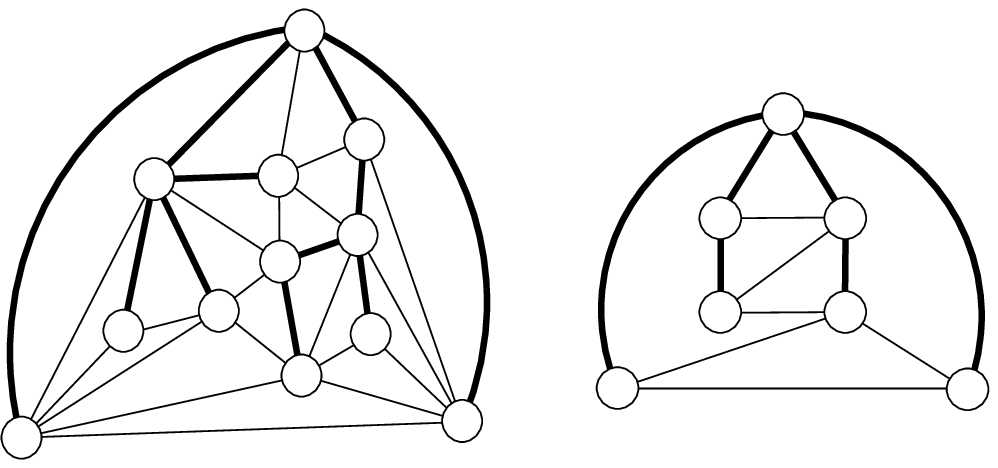}
\end{center}
\caption{(a) The tree rooted at node 1, consisting of the thick edges,
is not an orderly spanning tree of the plane graph. (b) A triconnected
plane graph $H$, where the thick edges form an orderly spanning tree
$T$, rooted at node 1, of $H$. The counterclockwise preordering of $T$
is not a canonical ordering of $H$.}
\label{fig:tricon}
\end{figure}

Let $T$ be a rooted spanning tree of a connected plane graph $H$. Two
distinct nodes of $H$ are {\em unrelated} with respect to $T$ if
neither of them is an ancestor of the other in $T$. An edge $e$ of $H$
is {\em unrelated} with respect to $T$ if the endpoints of $e$ are
unrelated with respect to $T$.
Let $v_1,v_2,\ldots,v_n$ be the counterclockwise preordering of the
nodes in $T$. A node $v_i$ is {\em orderly} in $H$ with respect to $T$
if the neighbors of $v_i$ in $H$ form the following four blocks of $H$ with respect to $T$ in
counterclockwise order around $v_i$:
\begin{enumerate}[\qquad$B_1{(v_i)}$:]
\item the parent of $v_i$ in $T$;
\item the nodes $v_j$ with $j<i$ that are unrelated to $v_i$ with respect
to $T$; 
\item the children of $v_i$ in $T$; and
\item the nodes $v_j$ with $j>i$ that are unrelated to $v_i$ with respect
to $T$, 
\end{enumerate}
where each block could be empty. $T$ is an {\em orderly spanning tree}
of $H$ if (i) $v_1$ is on the contour of $H$, and (ii) each node $v_i$
is orderly in $H$ with respect to $T$.  Clearly, if $T$ is an orderly
spanning tree of $H$, then each incident edge of $v_1$ in $H$ belongs
to $T$.  An example of orderly spanning tree is given in
Figure~\ref{fig:vibi-real}(a). Figure~\ref{fig:tricon}(a) provides a
negative example of orderly spanning tree, where nodes 1, 3, 8, and
10 are not orderly in $H$ with respect to $T$.

\begin{figure}
\begin{center}
\input{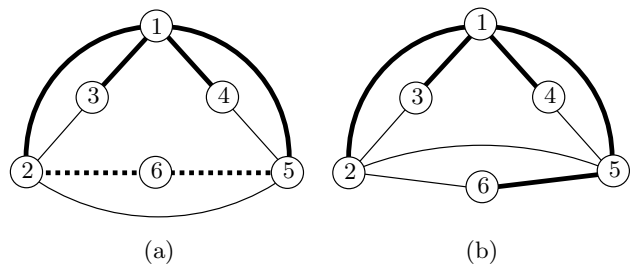}
\end{center}
\caption{(a) A plane graph $H$ that has no
orderly spanning trees. (b) A different planar embedding of $H$ that
admits an orderly spanning tree rooted at node 1, consisting of the
thick edges.}
\label{fig:embed}
\end{figure}

Not every connected plane graph admits an orderly spanning tree.
However, as to be shown in this section, there always exists a planar
embedding for any given planar graph that admits an orderly spanning
tree.  For example, consider the plane graph $H$ in
Figure~\ref{fig:embed}(a).  Assume for a contradiction that $H$ admits
an orderly spanning tree $T$ rooted at node 1. Observe that the thick
edges must be in $T$, and thus the thin edges cannot be in
$T$. Clearly, $T$ contains exactly one of the dashed edges. In either
case, however, the parent of node 6 in $T$ is not orderly in $H$ with
respect to $T$, thereby, contradicting the assumption that $T$ is an
orderly spanning tree rooted at node 1. Since $H$ is rotationally
symmetric, $H$ admits no orderly spanning trees.  If we change the
planar embedding of $H$ by moving edge $(2,5)$ to the interior of $H$,
as shown in Figure~\ref{fig:embed}(b), then the new plane graph has an
orderly spanning tree rooted at node 1 consisting of the thick edges.

We say that $(H,T)$ is an {\em orderly pair} of a connected planar
graph $G$ with respect to $r$ if (i) $H$ is a plane graph of $G$, and
(ii) $T$ rooted at $r$ is an orderly spanning tree of $H$.  The
concept of orderly pair originates from that of canonical spanning
tree of triconnected plane graphs, introduced by Chuang et
al.~\cite{ChuangGHKL98}. If a plane graph $H$ is triconnected, then an
orderly spanning tree of $H$ is precisely a canonical spanning tree of
$H$.  One the one hand, given a canonical ordering of $H$, obtainable
in linear time~\cite{Kant96}, it takes linear time to compute for $H$
an orderly spanning tree $T$ whose counterclockwise preordering is the
given canonical ordering of $H$~\cite{ChuangGHKL98}.  On the other
hand, as shown in Figure~\ref{fig:tricon}(b), the counterclockwise
preordering of an orderly spanning tree for $H$ may not be a canonical
ordering of $H$.  If $H$ is a plane triangulation, however, then it is
not difficult to verify that the counterclockwise preordering of any
orderly spanning tree of $H$ is a canonical ordering of $H$.

\subsection{The orderly-pair algorithm}
This subsection shows how to compute an orderly pair for any planar
graph in linear time.  Without loss of generality, we may assume that
the input planar graph is already equipped with a planar embedding
represented by an adjacency list, where each node $v$ keeps a doubly
linked list, storing its neighbors in counterclockwise order around
$v$.  Moreover, two copies of an edge are cross-linked to each
other. Based upon this representation, both deleting an edge and
moving an edge to the interior of a face can be conducted in $O(1)$
time. Such a representation can be obtained as a by-product by running
the linear-time planarity-testing algorithm of Hopcroft and
Tarjan~\cite{Hoprcoft74}.

To describe the algorithm, we need some definitions for a 2-connected
plane graph $H$. If $v$ is an external node in $H$, then let
$\next{H,v}$ (respectively, $\prev{H,v}$) denote the external node of
$H$ that immediately succeeds (respectively, precedes) $v$ in
counterclockwise order around the contour of $H$.  For any two
distinct external nodes $r$ and $v$ of $H$, let $K_1(H,r,v)$
(respectively, $K_2(H,r,v)$) denote the sequence of the external nodes
of $H$ from $r$ to $v$ in counterclockwise (respectively, clockwise)
order around the contour of $H$. Clearly, we have $\prev{H,v}\in
K_1(H,r,v)$ and $\next{H,v}\in K_2(H,r,v)$. Let
$K(H,r)=K_1(H,r,\prev{H,r})$, i.e., the sequence of the external nodes
of $H$ from $r$ to $\prev{H,r}$ in counterclockwise order around the
contour of $H$.  For example, if $H$ is the plane graph shown in
Figure~\ref{fig:embed}(b), then we have $\next{H,2}=6$,
$\prev{H,2}=1$, $K_1(H,1,6)=(1,2,6)$, $K_2(H,1,6)=(1,5,6)$, and
$K(H,1)=(1,2,6,5)$.

The key component of our orderly-pair algorithm is the following
recursive subroutine $\block(G,r,v)$, where $G$ is a 2-connected plane
graph, and $r$ and $v$ are two distinct external nodes of $G$.

\paragraph{\bf Subroutine $\block(G,r,v)$}

\begin{figure}
\centerline{\input{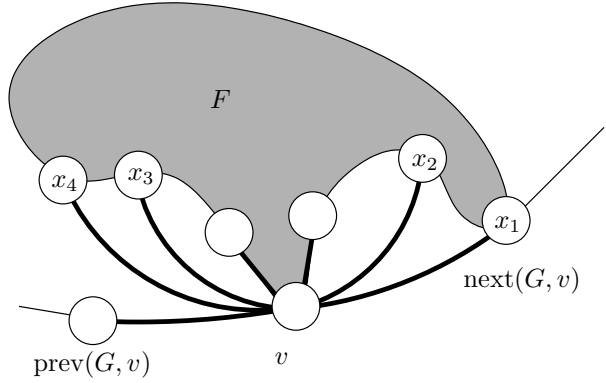}}
\caption{$F$ is an internal face of $G$ 
containing nodes $v$ and $x_i$, but not edge $(v,x_i)$ for each
$i\in\setof{1,2,3,4}$.}
\label{figure:mobile}
\end{figure}

\begin{enumerate}[Step 1.]

\item\label{step:1} If $G$ consists of a single edge $(r,v)$, then
return $(G,G)$; otherwise, perform Steps~\ref{step:2}--\ref{step:7}.

\item\label{step:2} Perform Step~\ref{step:2}.1 for each internal face
$F$ of $v$ in $G$ in clockwise order around $v$ starting from the one
containing $(v,\prev{G,v})$.

\begin{enumerate}[Step~\ref{step:2}.1.]
\item For any node $x$ in $F$ such that $(v,x)$ is an edge of $G$
preceding $F$ in counterclockwise order around $v$ starting from
$(v,\next{G,v})$, update the planar embedding of $G$ by flipping
$(v,x)$ into the interior of $F$.

{\em Remark.} For instance, if $v$ and $F$ are as shown in
Figure~\ref{figure:mobile}, then $(v,x_1)$ and $(v,x_2)$ will be
flipped into the interior of $F$ by Step~\ref{step:2}.1.
\end{enumerate}

\item\label{step:3}
Let $p$ be the neighbor of $v$ in $G$ closest to $r$ in
$K_2(G,r,v)$.

\item\label{step:4} Perform Step~\ref{step:4}.1 for each internal face
$F$ of $G$ that succeeds $(v,p)$ in counterclockwise order around $v$
starting from the one containing $(v,p)$:

\begin{enumerate}[Step~\ref{step:4}.1.]
\item For any node $x$ in $F$ such that $(v,x)$ is an edge of $G$
succeeding $F$ in counterclockwise order around $v$ starting from
$(v,\next{G,v})$, update the planar embedding of $G$ by flipping
$(v,x)$ into the interior of $F$.

{\em Remark.} For instance, if $v$ and $F$ are as shown in
Figure~\ref{figure:mobile}, then $(v,x_3)$ and $(v,x_4)$ will be
flipped into the interior of $F$ by Step~\ref{step:4}.1.
\end{enumerate}

\item\label{step:5}
Let $G'$ be the graph obtained by deleting all the incident edges of
$v$ in $G$, except for $(v,p)$.  Compute the 2-connected components of
$G'$ by traversing the segment of the contour of $G'$ from
$\prev{G,v}$ to $\next{G,v}$ in counterclockwise order around the
counter of $G'$.

{\em Remark.} Since $G$ is 2-connected, we know that all 2-connected
components of $G'$ are external to one another. Therefore, the above
traversal on part of the contour will suffice.  Also, by definitions
of $G'$ and $p$, some 2-connected component of $G'$ consists of the
single edge $(v,p)$.

\item\label{step:6}
Compute $(H_i,T_i)=\block(G_i,r_i,v_i)$ for each 2-connected
component $G_i$ of $G'$, where $r_i$ is the node of $G_i$ 
closest to $r$ in $G'$, and $v_i$ is defined as follows:
\begin{enumerate}[{Case} 1:]
\item $G_i=(v,p)$. Let $v_i=v$.
\item $G_i$ and $\prev{G,v}$ are on the same side of $(v,p)$ in $G$.
Let $S$ consist of the nodes in both
$K_1(G_i,\next{G_i,r_i},\prev{G_i,r_i})$ and $K_1(G,r,v)$. If $S$ is
empty, then let $v_i=\next{G_i,r_i}$.  Otherwise, let $v_i$ be the
last node of $S$ in counterclockwise order around the contour of $G_i$.

\item $G_i$ and $\next{G,v}$ are on the same side of $(v,p)$ in $G$.
Let $S$ consist of the nodes in both
$K_1(G_i,\next{G_i,r_i},\prev{G_i,r_i})$ and $K_2(G,r,v)$. If $S$ is
empty, then let $v_i=\prev{G_i,r_i}$. Otherwise, let $v_i$ be the
first node of $S$ in counterclockwise order around the contour of
$G_i$.
\end{enumerate}

\item\label{step:7}
Return $(H,T)$, where $H$ is obtained from $G$ by replacing each
$G_i$ with $H_i$, and $T$ is the union of all $T_i$.
\end{enumerate}

\begin{figure}
\centerline{\input{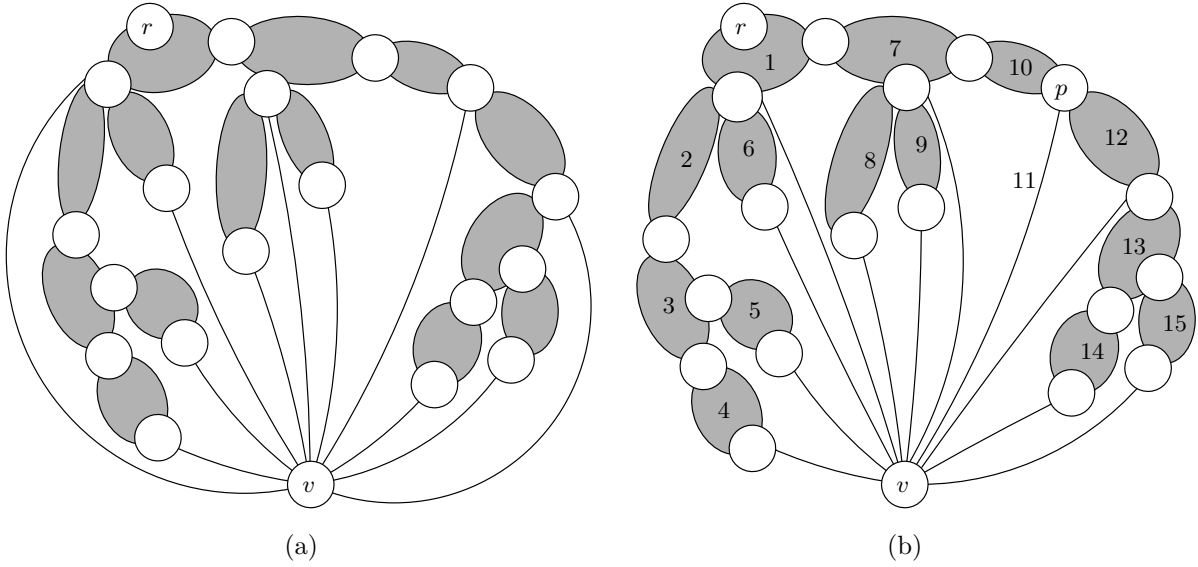}}
\caption{(a) A 2-connected plane graph $G$, where each gray ellipse is a
2-connected component of $G-\setof{v}$. (b) The plane graph $G$ at the
completion of performing Steps~\ref{step:1}--\ref{step:4} of
$\block(G,r,v)$.}
\label{figure:demo}
\end{figure}

An illustration of $\block(G,r,v)$ is given in
Figure~\ref{figure:demo}.  Let $G$ be the 2-connected plane graph
shown in Figure~\ref{figure:demo}(a).  At the completion of
Step~\ref{step:4}, the resulting $G$ and $p$ are as shown in
Figure~\ref{figure:demo}(b), where the gray ellipse with label $i$ is
the $i$-th 2-connected component $G_i$ of $G'$.  Note that $(v,p)$ is
also a 2-connected component of $G'$.  One can verify that after
Step~\ref{step:6} we have $r_1=r$, $r_2=r_6$, $r_8=r_9$,
$r_{11}=r_{12}=p$, and $v_{11}=v$.  For the 2-connected components
lying on the same side of $(v,p)$ with $\prev{G,v}$, we have
$v_1=r_2$, $v_2=r_3$, $v_3=r_4$, $v_4=\prev{G,v}$, and
$v_i=\next{G_i,r_i}$ for each $i\in\setof{5,6,\ldots,10}$. For the
2-connected components lying on the same side of $(v,p)$ with
$\next{G,v}$, we have $v_{12}=r_{13}$, $v_{13}=r_{15}$,
$v_{14}=\prev{G_{14},r_{14}}$, and $v_{15}=\next{G,v}$.

\begin{lemma}
\label{lemma:correct}
If $r$ and $v$ are two distinct external nodes of a 2-connected plane
graph $G$, then $\block(G,r,v)$ outputs an orderly pair of $G$ with
respect to $r$.
\end{lemma}
\begin{proof}
Let $(H,T)$ be the output of $\block(G,r,v)$.  We prove the following
statements with respect to $G$, $H$, $T$, $r$, and $v$ by induction on
the number of edges in $G$:
\begin{enumerate}

\item
\label{correct:a1}
Each external node of $G$ remains external in $H$. Moreover, $K(G,r)$
is a subsequence of $K(H,r)$.

\item
\label{correct:a2} For each neighbor $x$ of $v$ in $H$ other than $p$,
if $x$ and $\prev{H,v}$ (respectively, $\next{H,v}$) 
are on the same side of $(v,p)$ in $H$, then 
$(v,x)$ is on the first
(respectively, last) internal face of $H$ containing $v$ and $x$ in
counterclockwise order around $v$ starting from the one containing
$(v,\next{H,v})$. 

\item
\label{correct:a3}
$T$ rooted at $r$ is a spanning tree of $H$ such that
exactly one of the following conditions holds for each node $u$ in
$K_1(H,r,v)$ (respectively, $K_2(H,r,v)$): (i) $u$ is a leaf of $T$;
and (ii) $\next{H,u}$ (respectively, $\prev{H,u}$) is the
lowest-indexed (respectively, highest-indexed) child of $u$ in $T$.

\item 
\label{correct:a4}
$H$ is a plane graph of $G$.

\item 
\label{correct:a5}
$T$ rooted at $r$ is an orderly spanning tree of $H$.
\end{enumerate}
Statements~\ref{correct:a4} and~\ref{correct:a5} clearly suffice, but
we need the other statements to enable the induction step.  When $G$
consists of a single edge $(r,v)$, by Step~\ref{step:1} we have
$H=T=G$. It is not difficult to see the inductive basis of each
statement holds.  For brevity, let Statement~$j$($i$) stand for
Statement~$j$ with respect to $G_i$, $H_i$, $T_i$, $r_i$, and
$v_i$. Suppose $G'$ consists of $k$ 2-connected components.  By
Step~\ref{step:6}, we have $r_i\ne v_i$ for each $i$.  It follows from
the inductive hypothesis that Statement~$j$($i$) holds for each
$i\in\setof{1,2,\ldots,k}$ and $j\in\setof{1,2,\ldots,5}$.  The rest
of the proof shows the induction step.

\paragraph{Statement~\ref{correct:a1}} 
Observe that throughout the execution of $\block(G,r,v)$, without
accounting for its subsequent subroutine calls to $\block$, the
embedding of $G$ changes only by flipping edges into the interior of
internal faces of $G$ in Steps~\ref{step:2} and~\ref{step:4}. Thus,
based on how $H$ is obtained from $G$ in Step~\ref{step:7}, it follows
from Statement~\ref{correct:a1}($i$) for all
$i\in\setof{1,2,\ldots,k}$ that the statement holds.

\paragraph{Statement~\ref{correct:a2}} 
Let Statement~\ref{correct:a2}' stand for the statement obtained from
Statement~\ref{correct:a2} by replacing each $H$ with an $G$.  By
Steps~\ref{step:2} and~\ref{step:4}, one can easily verify that the
$G$ at the completion Step~\ref{step:4} satisfies
Statement~\ref{correct:a2}'. From Statement~\ref{correct:a2}
and how $H$ is obtained from $G$ in Step~\ref{step:7}, we know that
the relative order among the incident edges of $v$ and the faces
containing $v$ remains the same in $G$ and $H$. Therefore the
statement follows from Statement~\ref{correct:a2}'.

\paragraph{Statement~\ref{correct:a3}}
For each $i\in\setof{1,2,\ldots,k}$, Statement~\ref{correct:a3}($i$)
implies that $T_i$ is a spanning tree of $H_i$. Since
$H_1,H_2,\ldots,H_k$ are edge disjoint, and each node of $H$ belongs
to some $H_i$, we know that $T$, the union of all $T_i$, is a spanning
tree of $H$.  Since $v$ is a leaf of $T$, clearly the required
property holds for $v$.  Let $x$ be an external node of $H$ other than
$v$.  If $(x,v)$ is not an external edge of $H$ belonging to $H-T$,
then the required property for $x$ follows from the property of $x$
guaranteed by Statement~\ref{correct:a3}($i$) for each index $i$ with
$x\in H_i$.  Otherwise, by Statement~\ref{correct:a2}, $x$ is either
$\prev{H,v}$ or $\next{H,v}$. Let $H_j$ be the 2-connected component
of $H'$ containing $x$. We have $v_j=x$. By
Statement~\ref{correct:a3}($j$), $x$ is a leaf of $T_j$. Clearly, $x$
is also a leaf of $T$, so the required property holds for $x$.

\paragraph{Statement~\ref{correct:a4}}
Observe that Steps~\ref{step:2} and~\ref{step:4} flip an edge $(v,x)$
into the interior of $F$ only if $F$ contains both $v$ and
$x$. Therefore, the equipped embedding of $G$ at the completion of
Step~\ref{step:4} is still planar.  According to how $H$ is obtained
from $G$ in Step~\ref{step:7}, the statement follows from
Statements~\ref{correct:a1} and~\ref{correct:a4}($i$) for all indices
$i\in\setof{1,2,\ldots,k}$.

\paragraph{Statement~\ref{correct:a5}}
Clearly, each neighbor of $r$ in $H$ is a child of $r$ in $T$, hence
$r$ is orderly in $H$ with respect to $T$.  The rest of the proof
shows that each node $x$ other than $r$ is orderly in $H$ with respect
to $T$.  Let $H'$ be the graph obtained from $H$ by deleting each
incident edge of $v$ in $H-T$. Clearly, $H'$ is a plane graph of $G'$,
and each $H_i$ is a 2-connected component of $H'$.  Let $I_x$ consist
of the indices $i$ with $x\in H_i$. By $x\ne r$, one can verify that
there is an index $j$ in $I_x$ such that $x\ne r_j$ and $x=r_i$ for
each index $i\in I_x -\setof{j}$.

We first show that if $(v,x)$ is an edge of $H-H'$, then $(v,x)$ is
unrelated with respect to $T$.  If the index of $x$ is higher than
that of $v$, then by the fact that $v$ is a leaf in $T$, we know that
$(v,x)$ is unrelated. As for the case with the index of $x$ lower than
than that of $v$, let us assume for a contradiction that $x$ is an
ancestor of $v$ in $T$.  By $(v,x)\in H-T$ and the fact that $p$ is
the parent of $v$ in $T$, we know that $x$ is also an ancestor of $p$
in $T$.  Let $P$ be the path of $T$ between $r$ and $p$. Clearly,
$x\in P$. Let $y$ be the node of $H_{j}$ closest to $p$ in $P$. It is
not difficult to see $y\in K_2(H,r,p)$ and $y\in
K_2(H_{j},r_{j},x)$. By $(v,p)\in T$, $(v,x)\in H-T$, and $y\in
K_2(H,r,p)$, we know $y\ne x$. Otherwise, $x$ would have been a
neighbor of $v$ in $H$ closer to $r$ than $p$ in
$K_2(H,r,v)$, thereby, contradicting the choice of $p$ by
Step~\ref{step:3}.  By $y\ne x$, $x$ is not a leaf of $T_{j}$. Let
$z=\prev{H_{j},x}$. By Statement~\ref{correct:a2}($j$), we know that
$z$ is a child of $x$ in $T_{j}$.  By $(v,x)\not\in T$, we have $x\ne
r_{j}$.  By $x\ne r_{j}$ and $y\in K_2(H_{j},r_{j},x)$, we know that
$y$ and $z$ are on different sides of the path of $T_{j}$ between
$r_{j}$ and $x$ in $H_{j}$, thereby, contradicting the fact that $z$
is the highest-indexed child of $x$ in $T_{j}$.

We then show that $x$ is orderly in $H'$ with respect to $T$.  If
$|I_x|=1$, then the orderly pattern of $x$ in $H'$ with respect to $T$
follows immediately from that in $H_{j}$ with respect to $T_{j}$,
which is ensured by Statement~\ref{correct:a5}($j$). When $|I_x|\geq
2$, by Statement~\ref{correct:a5}($i$) for all $i\in I_x-\setof{j}$,
each neighbor of $x$ in $\bigcup_{i\in I_x-\setof{j}}H_{i}$ is a child
of $x$ in $T$. It follows from Statement~\ref{correct:a3}($j$) that
all children of $x$ in $T$ are consecutive in $H'$ around $x$.  Since
$x$ is orderly in $H_{j}$ with respect to $T_{j}$, one can see that
$x$ is orderly in $H'$ with respect to $T$.

Since $v$ is a leaf of $T$, we know that $v$ is orderly in $H$ with
respect to $T$.  It remains to show that each neighbor $x$ of $v$ in
$H-H'$ is orderly in $H$ with respect to $T$.  Let $z_1$
(respectively, $z_2$) be the neighbor of $x$ that precedes
(respectively, succeeds) $v$ in counterclockwise order around $x$.  It
suffices to show that if the index of $x$ is lower (respectively,
higher) than that of $v$, then $z_2$ (respectively, $z_1$) belongs to
$B_1(x)$ or $B_4(x)$ (respectively, $B_1(x)$ or $B_2(x)$) of $H'$ with
respect to $T$ as follows: If the index of $x$ is lower than that of
$v$, then we know $z_2=\next{H_{j},x}$ by Statement~\ref{correct:a2}.
By Step~\ref{step:6}, one can verify that $x$ belongs to
$K_2(H_{j},r_{j},v_{j})$. By Statement~\ref{correct:a3}, we have that
$z_2$ belongs to either $B_1(x)$ or $B_4(x)$ of $H'$ with respect to
$T$.  If the index of $x$ is higher than that of $v$, then we know
$z_1=\prev{H_{j},x}$ from Statement~\ref{correct:a2}. By
Step~\ref{step:6}, one can verify that $x$ belongs to
$K_1(H_{j},r_{j},v_{j})$. From Statement~\ref{correct:a3}, we have
that $z_1$ belongs to either $B_1(x)$ or $B_2(x)$ of $H'$ with respect
to $T$.
\end{proof}

\begin{lemma}
\label{lemma:time}
If $r$ and $v$ are two distinct external nodes of an $n$-node
2-connected plane graph $G$, then $\block(G,r,v)$ runs in $O(n)$ time.
\end{lemma}
\begin{proof}
The execution of $\block(G,r,v)$ consists of a sequence of subroutine
calls to $\block$. One can see that each node of $G$ can be the
parameter $v$ for no more than two subroutine calls to $\block$ ---
one with $G\ne (r,v)$ and the other with $G=(r,v)$. Clearly, if
$G=(r,v)$, then the subroutine call $\block(G,r,v)$ runs in $O(1)$
time.  Let $\ell$ be the number of subroutine calls to $\block(G,r,v)$
with $G\ne(r,v)$.  For each $j \in\setof{1,2,\ldots,\ell}$, let
$\block(G_j,r_j,v_j)$ be the $j$-th subroutine call to $\block$ with
$G_j\ne (r_j,v_j)$ throughout the execution of $\block(G,r,v)$, where
$G_1=G$, $r_1=r$, and $v_1=v$.  Clearly, $v_j\ne v_{j'}$ for any two
distinct indices $j$ and $j'$, thereby, implying $\ell \leq n$.  Let
$E_j$ consist of the edges of $G$ belonging to the boundary of some
internal face of $G_j$ which contains $v_j$. Let $t_j$ be the time
required by $\block(G_j,r_j,v_j)$, without accounting for that
required by its subsequent subroutine calls to $\block$. Clearly,
$t_j=O(|E_j|)$ holds for each $j$. It is not difficult to implement
the algorithm $\block$ such that the running time of $\block(G,r,v)$
is dominated by $\sum_{j=1}^\ell t_j =
\sum_{j=1}^\ell O(|E_j|)$.  Since $G$ has $O(n)$ edges, it suffices to
show as follows that any edge $(x,y)$ of $G$ belongs to no more than
two of the sets $E_1,E_2,\ldots,E_\ell$: Let $j_1$ be the smallest
index $j$ with $(x,y)\in E_j$. If $v_{j_1}\in\setof{x,y}$, then $j_1$
is also the largest index $j$ with $(x,y)\in E_j$. It remains to
consider the case $v_{j_1}\not\in\setof{x,y}$. Let $j_2$ be the
smallest index $j$ with $j>j_1$ and $(x,y)\in E_{j}$. By definition of
$\block$, it can be verified that $(x,y)$ has to be on the contour of
$G_{j_2}$, implying $v_{j_2}\in\setof{x,y}$. Therefore, $j_2$ is the
largest index $j$ with $(x,y)\in E_{j}$.
\end{proof}

Finally, we have the next theorem.
\begin{theorem}\label{theorem:tree}
It takes $O(n)$ time to compute an orderly pair for an $n$-node
connected planar graph.
\end{theorem}
\begin{proof} 
Let $G$ be a plane graph of the input $n$-node planar graph. Let
$G_1,G_2,\ldots,G_k$ be the 2-connected of $G$. Let $r$ be an external
node of $G_1$. For each $i$, let $r_i$ be the node of $G_i$ closest to
$r$ in $G$. Clearly, $r_1=r$. Also, for each
$i\in\setof{2,3,\ldots,k}$, it is not difficult to see that $r_i$ is
an external node of $G_i$, and that $G-\setof{r_i}$ is disconnected.
For each $i\in\setof{1,2,\ldots,k}$, let
$(H_i,T_i)=\block(G_i,r_i,\next{G_i,r_i})$.  Let $T$ be the union of
$T_1,T_2,\ldots,T_k$ rooted at $r$.  Let $H$ be the union of
$H_1,H_2,\ldots,H_k$ such that, for each $i\in\setof{1,2,\ldots,k}$,
the children of $r_i$ in $T$ are consecutive in the counterclockwise
neighbor sequence of $r_1$ in $H$. By Lemma~\ref{lemma:correct}, one
can verify that $H$ is a well-defined plane graph of $G$, and that $T$
is an orderly spanning tree of $H$. By Lemma~\ref{lemma:time}, it is
not difficult to see that both $H$ and $T$ are derivable in $O(n)$
time.
\end{proof}

\section{Realizers for plane triangulations}
\label{section:realizer} 
This section provides a new linear-time algorithm for computing a
realizer for any $n$-node plane triangulation $G$. As defined by
Schnyder~\cite{Schnyder90,Schnyder89}, $(T_1,T_2,T_n)$ is a {\em
realizer} of $G$ if
\begin{itemize}
\item the internal edges of $G$ are partitioned into three edge-disjoint
trees $T_1$, $T_2$, and $T_n$, each rooted at a distinct external
node of $G$; and
\item the neighbors of each internal node $v$ of $G$ form six blocks
$U_1$, $D_n$, $U_2$, $D_1$, $U_n$, and $D_2$ in counterclockwise order
around $v$, where for each $j\in\setof{1,2,n}$, $U_j$ (respectively,
$D_j$) consists of the parent (respectively, children) of $v$ in
$T_j$.
\end{itemize}
A realizer of the plane triangulation in
Figure~\ref{fig:vibi-real}(a) is shown in
Figure~\ref{fig:vibi-real}(c).

\begin{lemma}\label{lemma:realizer}
Given an orderly spanning tree of $G$, a realizer of $G$ is computable
in $O(n)$ time.
\end{lemma}
\begin{proof}
Let $T$ be the given orderly spanning tree of $G$. Let
$v_1,v_2,\ldots,v_n$ be the counterclockwise preordering of $T$, where
$v_1$, $v_2$, and $v_n$ are the external nodes of $G$ in
counterclockwise order. Clearly, $(v_1,v_2)$ and $(v_1,v_n)$ must be
in $T$. Since $G$ is a plane triangulation, and the edge of $G-T$ is
unrelated with respect to $T$, we know that both $B_2(v_i)$ and
$B_4(v_i)$ are nonempty for each $3\leq i \leq n-1$. Let $p_i$
(respectively, $q_i$) be the index of the last (respectively, first)
node in $B_2(v_i)$ (respectively, $B_4(v_i)$) in counterclockwise
order around $v_i$. Let $T_1$ be obtained from $T$ by deleting
$(v_1,v_2)$ and $(v_1,v_n)$. Let $T_2=\set{(v_i,v_{p_i})}{3\leq i\leq
n-1}$ and $T_n=\set{(v_i,v_{q_i})}{3\leq i\leq n-1}$.  An example is
shown in Figure~\ref{fig:vibi-real}(c).  Clearly, $p_i< i< q_i$ holds
for each $3\leq i\leq n-1$, implying that both $T_2$ and $T_n$ are
acyclic. Since $G$ is a plane triangulation, exactly one of the
equalities $i=p_j$ and $j=q_i$ holds for each edge $(v_i, v_j)\in G-T$
with $i<j$. It follows that each internal edge of $G$ belongs to
exactly one of $T_1$, $T_2$, and $T_n$. By definitions of $p_i$ and
$q_i$, one can verify that the neighbors of each internal node $v_i$
of $G$ indeed form the required pattern for $(T_1,T_2,T_n)$ as a
realizer of $G$. Since it takes $O(1)$ time to determine each $p_i$
and $q_i$, the lemma is proved.
\end{proof}

\begin{theorem}[see also~\cite{Schnyder90,Schnyder89}]\label{theorem:realizer}
A realizer of any plane triangulation is derivable in linear time.
\end{theorem}
\begin{proof}
Straightforward by Theorem~\ref{theorem:tree} and
Lemma~\ref{lemma:realizer}.
\end{proof}

\section{2-visibility drawings for plane graphs}
\label{section:drawing} 
This section shows how to obtain in $O(n)$ time an $(n-1)\times
\floor{\frac{2n+1}{3}}$ 2-visibility drawing for any $n$-node plane
graph $G$.  For calculating the area of a 2-visibility drawing, we
follow the convention~\cite{FoessmeierKK97}, stating that the corner
coordinates of each rectangle are integers, and that each rectangle is
no smaller than $1\times 1$.  For example, the area of the
2-visibility drawing shown in Figure~\ref{fig:vibi-real}(b) is
$9\times 8$. Let $G'$ be a plane triangulation obtained by
triangulating $G$. It is clear that any 2-visibility drawing of $G'$
is also a 2-visibility drawing of $G$. The rest of the section assumes
that $G$ is a plane triangulation.

\begin{figure}
\centerline{\input{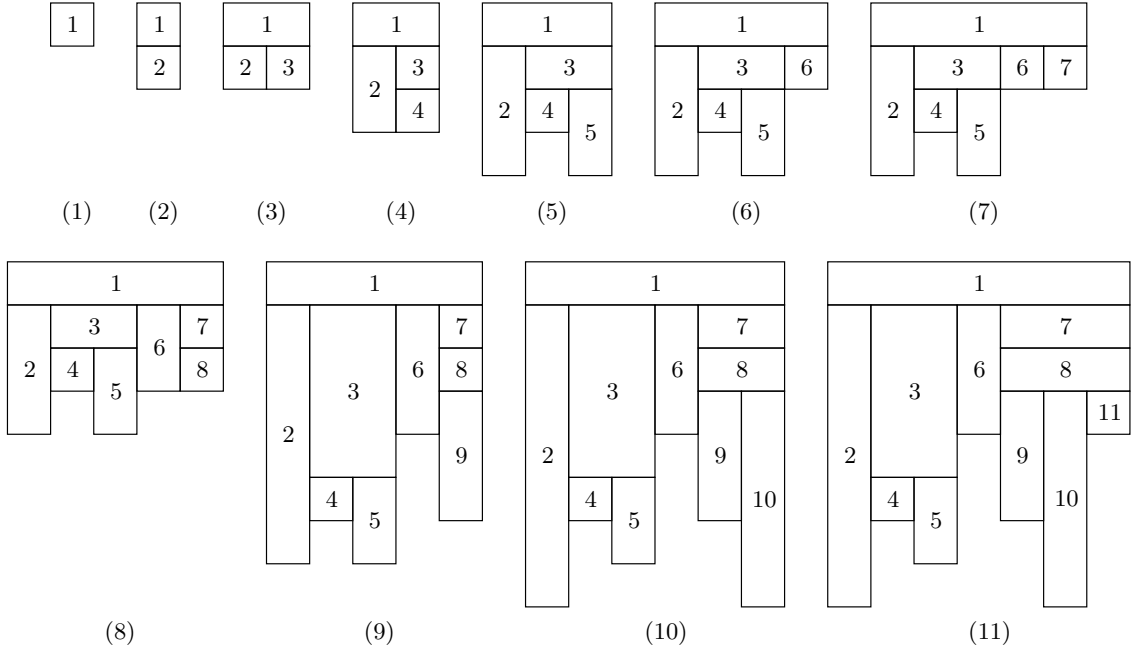}}
\caption{An illustration of the intermediate steps of $\draw(G,T)$,
where $G$ and $T$ are as shown in Figure~\ref{fig:vibi-real}(a).}
\label{fig:drawing}
\end{figure}

Let $T$ be an orderly spanning tree of $G$. Let $v_1,v_2,\ldots,v_n$
be the counterclockwise preordering of the nodes in $T$.  Our
algorithm $\draw(G,T)$ consists of $n$ iterations, where the
$i$-th iteration performs the following steps:
\begin{enumerate}[Step 1.]
\item If $i\ne 1$ and $v_i$ is not the first child of its parent in
$T$, then lengthen each ancestor of $v_i$ in $T$ to the right by one
unit.

\item Draw $v_i$ as a unit square beneath the parent of $v_i$ in $T$
such that $v_i$ and all ancestors of $v_i$ in $T$ align along the
right boundary. Clearly, $v_i$ is vertically visible to its parent in
$T$.

\item Lengthen downward $v_i$ and each neighbor $v_j$ of $v_i$ in $G$
with $j<i$, if necessary, so that $v_i$ and $v_j$ are horizontally
visible to each other.
\end{enumerate}
If $G$ and $T$ are as shown in Figure~\ref{fig:vibi-real}(a), then
the intermediate (respectively, resulting) drawing obtained by 
$\draw(G,T)$ is shown in Figure~\ref{fig:drawing} (respectively,
Figure~\ref{fig:vibi-real}(b)).

\begin{lemma}
\label{lemma:visibility-1} 
The algorithm $\draw(G,T)$ obtains an $x\times y$ 2-visibility drawing
of $G$ with $x\leq n-1$ and that $y$ equals the number of leaves in
$T$.
\end{lemma}
\begin{proof}
Since $T$ is an orderly spanning tree of $G$, it is clear that
$\draw(G,T)$ is well defined, and that the output of $\draw(G,T)$ is
indeed a 2-visibility drawing of $G$ with width equal to the number of
leaves in $T$. The rest of the proof shows that the height of the
output of $\draw(G,T)$ is at most $n-1$.  For any two distinct edges
$e$ and $e'$ in $G-T$, we say that $e$ {\em encloses} $e'$ if $e'$ is
enclosed by the cycle consisting of $e$ and the path of $T$ between
the endpoints of $e$.  Let $\hat{e}=(v_2,v_n)$.  Clearly, $\hat{e}$
encloses all the other edges in $G-T$.  For each edge $e$ in $G-T$, if
$e$ does not enclose any other edge in $G-T$, then let $\level{e}=1$;
otherwise, let $\level{e}$ be one plus the maximum of $\level{e'}$
over all the edges $e'$ in $G-T$ that are enclosed by $e$.  One can
easily verify that the height of the output of $\draw(G,T)$ is at most
$1+\level{\hat{e}}$. It remains to show $\level{\hat{e}}\leq n-2$ as
follows: Assume for a contradiction that $e_1,e_2,\ldots,e_{n-1}$ is a
sequence of edges in $G-T$ such that $e_i$ encloses
$e_1,e_2,\ldots,e_{i-1}$ for each $i\in\setof{2,3,\ldots,n-1}$. For
each $i\in\setof{1,2,\ldots,n-1}$, let $X_i$ consist of the endpoints
of $e_i,e_{i+1},\ldots,e_{n-1}$. Clearly, for each
$i\in\setof{1,2,\ldots,n-2}$, there must be an endpoint of $e_i$ that
is not in $X_{i+1}$. Therefore, $X_1$ contains at least $n$ distinct
nodes. Since $T$ is an orderly spanning tree of $G$, $v_1$ is not
incident to any edges of $G-T$. Therefore, $v_1\not\in X_1$, and
thereby, contradicting that $G$ has $n$ nodes.
\end{proof}

\begin{lemma}
\label{lemma:visibility-2}
It takes $O(n)$ time to compute an orderly spanning tree of $G$ with
$\floor{\frac{2n+1}{3}}$ or fewer leaves.
\end{lemma}

\begin{proof}
Let $v_1$, $v_2$, and $v_n$ be the external nodes of $G$ in
counterclockwise order around the contour of $G$. By
Theorem~\ref{theorem:realizer}, a realizer $(T'_1,T'_2,T'_n)$ of $G$,
where each $T'_i$ is rooted at $v_i$, can be obtained in $O(n)$ time.
Let $I=\setof{1,2,n}$.  For each $i\in I$, let $T_i=T'_i\cup
\setof{(v_i,v_{i_1}),(v_i,v_{i_2})}$, where
$\setof{i_1,i_2}=I-\setof{i}$.  Clearly, $T_1$, $T_2$, and $T_n$ are
three spanning trees of $G$ with $T_1\cup T_2\cup T_n=G$.  We first
show that each $T_i$ is an orderly spanning tree of $G$.  Since the
relation among $T_1$, $T_2$, and $T_n$ is rotationally symmetric, it
suffices to verify that each node is orderly with respect to $T_1$.
Let $v_1,v_2,\ldots,v_n$ be the counterclockwise preordering of $T_1$.
For each $i \in I$ and $j\in\setof{1,2,\ldots,n}$, let $P_{i,j}$ be
the path of $T_i$ between $v_i$ and $v_j$.  Note that $P_{1,j}$,
$P_{2,j}$, and $P_{n,j}$ are three edge-disjoint paths of $G$ that
intersect only at $v_j$. Clearly, if $v_j$ is not a leaf of $T_1$,
then the children of $v_j$ in $T_1$ are consecutive in $G$ in
counterclockwise order around $v_j$.  Therefore, to ensure that each
node is orderly with respect to $T_1$, is suffices to prove that each
edge of $G-T_1$ is unrelated with respect to $T_1$: If $v_{j'}$ were
an ancestor of $v_j$ that is also a neighbor of $v_j$ in $G-T_1$, then
$v_j$ and $v_{j'}$ would be on different sides of $P_{2,j''}\cup
P_{n,j''}$ in $G$, where $v_{j''}$ is the parent of $v_j$ in $T_1$,
thereby, contradicting the planarity of $G$.

It remains to show that $T_1$, $T_2$, or $T_n$ has at most
$\frac{2n+1}{3}$ leaves. For each $i\in I$, let $\leaf{i}$ consist of
the leaves of $T'_i$. Since the number of leaves in $T_i$ is precisely
$2+|\leaf{i}|$, it suffices to show $\sum_{i\in I}|\leaf{i}|\leq 2n-5$
as follows.  Let $v$ be a node in $\leaf{i}$.  Clearly, $v$ is
internal in $G$. For each $i\in I$, let $p_i(v)$ denote the parent of
$v$ in $T_i$.  Let $i_1$ and $i_2$ be the indices in $I-\setof{i}$.
Since $(T'_1,T'_2,T'_n)$ is a realizer of $G$, there is a unique
internal face $F_i(v)$ of $G$ containing $v$, $p_{i_1}(v)$, and
$p_{i_2}(v)$.  Clearly, $p_{i_1}(v)\not\in\leaf{i_1}$ and
$p_{i_2}(v)\not\in\leaf{i_2}$. It follows that $F_i(v)\ne
F_{i_1}(u_1)$ for any node $u_1$ in $\leaf{i_1}$, and that $F_i(v)\ne
F_{i_2}(u_2)$ for any node $u_2$ in $\leaf{i_2}$. Therefore,
$\sum_{i\in I}|\leaf{i}|$ is no more than the number of internal faces
of $G$, which is precisely $2n-5$ by Euler's formula.
\end{proof}

\begin{theorem}
An $(n-1)\times\floor{\frac{2n+1}{3}}$ 2-visibility drawing of any
$n$-node planar graph is computable in $O(n)$ time.
\end{theorem}
\begin{proof}
It is not difficult to see that the algorithm $\draw$ can be
implemented to run in linear time, so the theorem follows immediately
from Lemmas~\ref{lemma:visibility-1} and~\ref{lemma:visibility-2}.
\end{proof}

\section{Convenient encodings for planar graphs}
\label{section:data-structure} 
This section gives the best known convenient encodings for planar
graphs as an application of our orderly-pair algorithm.  We need some
notations to describe the data structures required by our convenient
encodings. Let $|S|$ denote the length of a string $S$.  Clearly, an
$S$ consisting of $t$ distinct symbols can be encoded in
$|S|\ceiling{\log_{2}t}$ bits.  For example, if $S$ consists of
parentheses and brackets, including open and close ones, then $S$ can
be encoded in $2|S|$ bits.  $S$ is {\em binary} if it consists of no
more than two distinct symbols. For each $1\leq i\leq j\leq |S|$, let
$S[i,j]$ be the length-$(j-i+1)$ substring of $S$ from the $i$-th
position to the $j$-th position. If $i>j$, then let $S[i,j]$ be the
empty string. Define $S[i]=S[i,i]$. $S[k]$ is {\em enclosed} by $S[i]$
and $S[j]$ in $S$ if $i<k<j$. Let $\select{S,i,\Box}$ be the position
of the $i$-th $\Box$ in $S$.  Let $\rank{S,k,\Box}$ be the number of
$\Box$'s before or at the $k$-th position of $S$. Clearly, if
$k=\select{S,i,\Box}$, then $i=\rank{S,k,\Box}$.

An {\em auxiliary string} $\chi$ of $S$ is a binary string with
$|\chi|=o(|S|)$ which is obtainable from $S$ in $O(|S|)$ time.

\begin{fact}[see~\cite{BCW90,Elias75}]
\label{fact:aux}
For any strings $S_1,S_2,\ldots,S_k$ with $k=O(1)$, there is an
auxiliary string $\aux_0$ such that, given the concatenation of
$\aux_0,S_1,S_2,\ldots,S_k$ as input, the index of the first symbol of
any given $S_i$ in the concatenation is computable in $O(1)$ time.
\end{fact}

Let $S_1+S_2+\cdots+S_k$ denote the concatenation of $\aux_0,
S_1,S_2,\ldots,S_k$ as in Fact~\ref{fact:aux}.

Suppose $S$ is a string of multiple types of parentheses. Let
$\rev{S}$ be the string $R$ such that $R[i]$ is the opposite type of
parenthesis $S[|S|+1-i]$. For example,
$\rev{\text{``$\rightp{}\leftp{}\rightp{}\rightb{}\rightp{}\leftb{}$''}}=
\text{``$\rightb{}\leftp{}\leftb{}\leftp{}\rightp{}\leftp{}$''}$.
For an open parenthesis $S[i]$ and a close one $S[j]$ of the same type
with $i<j$, the two {\it match} in $S$ if every parenthesis of the
same type that is enclosed by them matches one enclosed by them. $S$
is {\em balanced in type $k$} if every parenthesis of type $k$ in $S$
belongs to a matching parenthesis pair. $S$ is {\em balanced} if $S$
is empty or is balanced in all types of parentheses. Here are some
queries defined for a balanced $S$.  Let $\match{S,i}$ be the position
of the parenthesis in $S$ that matches $S[i]$.  Let
$\enclosek{k}{S,i_1,i_2}$ be the position pair $(j_1,j_2)$ of the
closest matching parenthesis pair of the $k$-th type that encloses
$S[i_1]$ and $S[i_2]$.

\begin{fact}[see~\cite{MR97,ChuangGHKL98}]
\label{fact:aux-string}
For any balanced string $S$ of $O(1)$ types of parentheses, there is
an auxiliary string $\aux_1(S)$ such that each of $\rank{S,i,\Box}$,
$\select{S,i,\Box}$, $\match{S,i}$, and $\enclosek{k}{S,i,j}$ can be
determined from $S+\aux_1(S)$ in $O(1)$ time.
\end{fact}

For a string $S$ of parentheses that may be unbalanced, we define
$\son{S,i}$ as follows.  For the case that $S[i]$ is an open
parenthesis of type $k$, let $S'$ be a string obtained from $S$ by
appending some close parentheses of type $k$, if necessary, such that
$S'[i]$ is matched in $S'$. Define $\son{S,i}$ to be the number of
indices $j$ satisfying $i< j\leq |S|$,
$\enclosek{k}{S',j,\match{S',j}}=(i,\match{S',i})$, and that $S[j]$ is
of type $k$.  For the case that $S[i]$ is closed, let
$\son{S,i}=\son{\rev{S},|S|+1-i}$. Therefore, if
\begin{eqnarray}\label{eq:S}
 S&=&\rm{\verb+(()[[[[[(](])[(]])[[)[[(])[[[(](](]]])[(]]])[[(])[)[)[(]]]]]))+},\\
  & &\rm{\verb+122.....3.4.4.5..5..3..6.6...7.8.9...9.A...A..B.B.8.7.C.....C1+}\nonumber
\end{eqnarray}
then we have $\son{S,1}=10$ and $\son{S,6}=4$. Clearly, if $S$ is
balanced, then $\son{S,i}$ is an even number for each $i$.  The next
lemma extends the set of queries supported in
Fact~\ref{fact:aux-string}.

\begin{lemma}\label{lemma:son}
For any balanced string $S$ of $O(1)$ types of parentheses, there is
an auxiliary string $\aux_2(S)$ such that $\son{S,i}$ can be computed
from $S+\aux_2(S)$ in $O(1)$ time.
\end{lemma}

\begin{proof}
Let $s=|S|$. Let $t$ be the number of distinct types of parentheses in
$S$.  Let $b$ be the smallest integer with $2t\leq 2^b$.  Clearly,
each symbol of $S$ can be encoded in $b$ bits.  By $t=O(1)$, we have
$b=O(1)$. Let $\ell=\floor{\frac{1}{2}\log_{2^b}s}$. Clearly, any
substring $S[i,j]$ with $j\leq i+\ell-1$ has $O(\sqrt{s})$ possible
distinct values.  Define tables $M_1$ and $M_2$ for $S$ by letting
$M_1[S[i,i+\ell-1]]=\son{S[i,i+\ell-1],1}$ and
$M_2[S[i,j]]=\son{\rev{S[i,j]},1}$, for any $i,j$ with $1\leq i\leq
j\leq i+\ell-1$. Clearly, $|M_1|=o(s)$ and $|M_2|=o(s)$.  For each
$k\in\setof{1,2,\ldots,t}$, define tables $M^k_3$ and $M^k_4$ as
follows. For each $i=1,2,\ldots,\ceiling{\frac{s}{\ell^2}}$, let
$M^k_3[i]=(j,\son{S,j})$, where $j$ is the largest index satisfying
$j\leq i\ell^2$, $\son{S,j}>\ell^2$, and that $S[j]$ is a close
parenthesis of type $k$. Let $M^k_4[i]=(j,\son{S,j})$, where $j$ is
the smallest index satisfying $j\geq i\ell^2$, $\son{S,j}>\ell^2$, and
that $S[j]$ is an open parenthesis of type $k$. Clearly,
$|M^k_3|=o(s)$ and $|M^k_4|=o(s)$.

An open $S[i]$ is {\em special} if (i) $\match{S,i}-i>\ell$, (ii)
$\son{S,i}\leq\ell^2$, and (iii) for each $S[j]$ with $j>i$ and
$S[j]=S[i]$, we have $j-i>\ell$ or $\match{S,i}-\match{S,j}>\ell$.  A
close $S[i]$ is {\em special} if $S[\match{S,i}]$ is special.  For
each $k\in\setof{1,2,\ldots,t}$, define tables $M^k_5$ and $M^k_6$ as
follows. For each $i\in\setof{1,2,\ldots,\ceiling{\frac{s}{\ell}}}$,
let $M^k_5[i]=(j,\son{S,j})$, where $j$ is the largest index with
$j\leq i\ell$ such that $S[j]$ is a special close parenthesis of type
$k$. Let $M^k_6[i]=(j,\son{S,j})$, where $j$ is the smallest index
with $j\geq i\ell$ such that $S[j]$ is a special open parenthesis of
type $k$. Clearly, $|M^k_5|=o(s)$ and $|M^k_6|=o(s)$.  Let
$\aux_2(S)=M_1+M_2+M^1_3+M^1_4+M^1_5+M^1_6+\cdots+M^t_3+M^t_4+M^t_5+M^t_6$.
It follows from $t=O(1)$ that $|\aux_2(S)|=o(s)$. Clearly, $\aux_2(S)$
can be derived from $S$ in $O(s)$ time.
\begin{figure}
\begin{tabbing}
\qquad\qquad\qquad\qquad\qquad\=\qquad\=\kill
\>function $\algo{\son{$S,i$}}$ \{\\
\>\>Step 1. let $k$, with $1\leq k\leq t$, be the type of $S[i]$;\\
\>\>Step 2. let $i_1=\min\setof{i,\match{S,i}}$;\\
\>\>Step 3. let $i_2=\match{S,i_1}$;\\
\>\>Step 4. let $(j,c)=M^k_3\left[\ceiling{\frac{i_2}{\ell^2}}\right]$; if $j=i_2$ then return $c$;\\
\>\>Step 5. let $(j,c)=M^k_4\left[\floor{\frac{i_1}{\ell^2}}\right]$; if $j=i_1$ then return $c$;\\
\>\>Step 6. let $(j,c)=M^k_5\left[\ceiling{\frac{i_2}{\ell}}\right]$; if $j=i_2$ then return $c$;\\
\>\>Step 7. let $(j,c)=M^k_6\left[\floor{\frac{i_1}{\ell}}\right]$; if $j=i_1$ then return $c$;\\
\>\>Step 8. let $j_1=i_1+\ell-1$;\\
\>\>Step 9. if $i_2-i_1\leq 2\ell$ then let $j_2=i_1+\ell$;\\
\>\>\qquad\qquad\qquad\qquad\qquad else let $j_2=i_2-\ell+1$;\\
\>\>Step 10. return $M_1[S[i_1,j_1]]$+$M_2[S[j_2,i_2]]$;\\
\>\}
\end{tabbing}
\caption{An $O(1)$-time algorithm that computes $\son{S,i}$.}
\label{figure:nextlevel}
\end{figure}

It remains to show that $\son{S,i}$ can be determined from $S$ and
$\aux_2(S)$ by the algorithm shown in Figure~\ref{figure:nextlevel},
which clearly runs in $O(1)$ time.  By definitions of
$M^k_3,M^k_4,M^k_5,M^k_6$, if a value $c$ is returned from Steps~4--7,
then $c=\son{S,i}$.

The rest of the proof assumes that Step~8 is executed. We first show
that $S[i]$ is not special and that $\son{S,i}\leq\ell^2$ holds.
Assume for a contradiction that $S[i]$ is special. By definitions of
$M^k_5$ and $M^k_6$, there exists an index $j$ such that (a)
$S[j]=S[i]$; (b) $S[j]$ and $S[\match{S,j}]$ encloses $S[i]$ and
$S[\match{S,i}]$; (c) $S[j]$ is special; (d) $1\leq |j-i|<\ell$; and
(e) $1\leq|\match{S,j}-\match{S,i}|<\ell$. By definition of special
parentheses, Condition~(c) contradicts Conditions~(d) and~(e). Assume
$\son{S,i}>\ell^2$ for contradictory purpose.  By definitions of
$M^k_3$ and $M^k_4$, there exists an index $j$ such that (a)
$S[j]=S[i]$; (b) $S[j]$ and $S[\match{S,j}]$ encloses $S[i]$ and
$S[\match{S,i}]$; (c) $\son{S,j}>\ell^2$; (d) $1\leq |j-i|<\ell^2$;
and (e) $1\leq|\match{S,j}-\match{S,i}|<\ell^2$. By Conditions~(d)
and~(e), we know
$\son{S,j}\leq\frac{1}{2}(|j-i|-1+|\match{S,j}-\match{S,i}|-1)+1<\ell^2$,
thereby, contradicting Condition~(c).

Now we are ready to argue that the algorithm correctly returns
$\son{S,i}$ in Step~10. By Steps~2 and~3, $S[i_1]$ is open and
$S[i_2]$ is closed. By Steps~8 and~9, we know $j_1<j_2$.  If
$i_2-i_1\leq 2\ell$, then 
\begin{eqnarray*}
M_1[S[i_1,j_1]]+M_2[S[j_2,i_2]] &=& 
\son{S[i_1,j_1],1}+\son{\rev{S[j_1+1,i_2]},1}\\
    &=&\son{S,i}.
\end{eqnarray*}
Now we assume $i_2-i_1>2\ell$.  Since $S[i_1]$ is not special and
$\son{S,i_1}\leq\ell^2$, by definition of special parentheses, there
exists an index $j'_1$ with $S[j'_1]=S[i_1]$, $0<j'_1-i_1\leq\ell$,
and $0<i_2-j'_2\leq\ell$, where $j'_2=\match{S,j'_1}$. Therefore, it
is not difficult to see
$M_1[S[i_1,j_1]]+M_2[S[j_2,i_2]]=\son{S[i_1,j_1],1}+\son{\rev{S[j_2,i_2]},1}=
\son{S[i_1,j'_1],1}+\son{\rev{S[j'_2,i_2]},1}=\son{S,i}$, thereby,
proving the lemma.
\end{proof}

A folklore encoding~\cite{HKL,MR97,ChuangGHKL98} $S$ of an $n$-node
simple rooted tree $T$ is a balanced string of $2n$ parentheses
representing a counterclockwise depth-first traversal of
$T$. Initially, an open (respectively, closed) parenthesis denotes a
descending (respectively, ascending) edge traversal. Then, this string
is enclosed by an additional matching parenthesis pair. For example,
the string in Equation~(\ref{eq:S_1}) is the folklore encoding for the
tree $T$ in Figure~\ref{fig:vibi-real}(a).  Let $v_i$ be the $i$-th
node in the counterclockwise depth-first traversal. Let $\rightp{i}$
be the close parenthesis of $S$ that matches $\leftp{i}$ in $S$.
Clearly, $v_i$ corresponds to $\leftp{i}$ and $\rightp{i}$ in that
$v_i$ is the parent of $v_j$ in $T$ if and only if $\leftp{i}$ and
$\rightp{i}$ form the closest pair of matching parentheses that
encloses $\leftp{j}$ and $\rightp{j}$. Also, the number of children of 
$v_i$ in $T$ is precisely $\son{S,\select{S,i,\leftp{}}}$, which is also
equal to $\son{S,\match{S,\select{S,i,\leftp{}}}}$.

Let $H$ be an $n$-node connected plane graph that may have multiple
edges but no self-loops. Let $T$ be a spanning tree of $H$ rooted at
$v_1$. Let $v_1v_2\cdots v_n$ be the counterclockwise preordering of
$T$.  Let $\text{degree}(i)$ be the number of edges incident to $v_i$
in $H$.  Let $\text{children}(i)$ be the number of children of $v_i$
in $T$. Let $\text{above}(i)$ (respectively, $\text{below}(i)$) be the
number of edges $(v_i,v_j)$ of $H$ such that $v_j$ is the parent
(respectively, a child) of $v_i$ in $T$.  Let $\text{low}(i)$
(respectively, $\text{high}(i)$) be the number of edges $(v_i,v_j)$ of
$H$ such that $j<i$ (respectively, $j>i$) and $v_j$ is neither the
parent nor a child of $v_i$ in $T$.  
Clearly,
$\text{degree}(i)=\text{above}(i)+\text{below}(i)+
\text{low}(i)+\text{high}(i)$. If $H$ has no multiple edges, then
$\text{below}(i)=\text{children}(i)$.  If $H$ and $T$ are as shown in
Figure~\ref{fig:vibi-real}(a), for instance, then $\text{above}(3)=1$,
$\text{below}(3)=\text{children}(3)=2$, $\text{low}(3)=1$, 
$\text{high}(3)=2$, and $\text{degree}(3)=6$.

The {\em $T$-code} of $H$ is a triple $(S_1,S_2,S_3)$ of binary
strings, where $S_1$, $S_2$, and $S_3$ are defined as follows:
\begin{itemize}
\item $S_1$ is the folklore encoding of $T$.

\item Let $p_i=\select{S_1,i,\leftp{}}$ and $q_i=\match{S_1,p_i}$.
$S_2$ has exactly $2n$ copies of $1$, in which $\text{low}(i)$
copies of $0$ immediately succeeds the $p_i$-th $1$, and
$\text{high}(i)$ copies of $0$ immediately succeeds the $q_i$-th
$1$.

\item $S_3$ has exactly $n$ copies of $1$, where
$\text{above}(i)+\text{below}(i)-\text{children}(i)-\delta_{i\geq
2}$ copies of $0$ immediately succeeds the $i$-th $1$.

\end{itemize}
For example, if $H$ and $T$ are as shown in
Figure~\ref{fig:vibi-real}(a), then
\begin{eqnarray}
   S_1 &=& \rm{\verb+(()(()())()((()()()))())+};\label{eq:S_1}\\
   S_2 &=& \rm{\verb+11100000101010100100100101000101010001010001001010101010000011+};\nonumber\\
   S_3 &=& \rm{\verb+111111111111+}.\nonumber
\end{eqnarray}
Clearly, we have
\begin{eqnarray*}
  |S_1|&=&2n;\\
  |S_2|&=&2n+\sum_{i=1}^{n}\left(\text{low}(i)+\text{high}(i)\right);\\
  |S_3|&=&\sum_{i=1}^{n}
          \left(\text{above}(i)+\text{below}(i)-\text{children}(i)\right)+1.
\end{eqnarray*}
Therefore, $|S_1|+|S_2|+|S_3|=2m+3n+2$. Moreover, if $H$ has no
multiple edges, then $|S_3|=n$ and thus, $|S_1|+|S_2|=2m+2n+2$.

The next theorem describes our convenient encodings.
\begin{theorem}\label{theorem:main}
Let $G$ be an input $n$-node $m$-edge planar graph having no
self-loops. If $G$ has $($respectively, has no$)$ multiple edges, then
$G$ has a convenient encoding, obtainable in $O(m+n)$ time, with
$2m+3n+o(m+n)$ $($respectively, $2m+2n+o(n)$$)$ bits.
\end{theorem}
\begin{proof} The techniques in the proof are mostly
adapted from~\cite{ChuangGHKL98}. We focus on the case that $G$ is
connected. It is not difficult to remove this restriction. By
Theorem~\ref{theorem:tree}, an orderly pair $(H,T)$ of $G$ can be
derived in $O(n)$ time.  Let $(S_1,S_2,S_3)$ be the $T$-code of $H$.
We prove that there exists an $o(m+n)$-bit string $\aux$, obtainable
in $O(m+n)$ time, such that $S_1+S_2+S_3+\aux$ is a convenient
encoding of $G$.  Clearly, if $G$ has no multiple edges, then $S_3$
consists of $n$ copies of $1$, and thus, $S_1+S_2+\aux$ will suffice.

If $p_i=\select{S_1,i,\leftp{}}$ and $q_i=\match{S_1,p_i}$, then
$\text{low}(i)=\select{S_2,p_i+1,1}-\select{S_2,p_i,1}-1$ and
$\text{high}(i)=\select{S_2,q_i+1,1}-\select{S_2,q_i,1}-1$.  Clearly,
we have $\text{children}(i)=\son{S_1,p_i}/2$. From definition of
$S_3$, we know $\text{above}(i)+\text{below}(i)-\text{children}(i)=
\select{S_3,i+1,1}-\select{S_3,i,1}-1+\delta_{i\geq 2}$. Let
$\aux'=\aux_1(S_1)+\aux_1(S_2)+\aux_1(S_3)+\aux_2(S_1)$. From
$\text{degree}(i)=\text{above}(i)+\text{below}(i)+\text{low}(i)+
\text{high}(i)$, Fact~\ref{fact:aux-string}, and
Lemma~\ref{lemma:son}, we determine that $\text{degree}(i)$ is
computable from $S_1+S_2+S_3+\aux'$ in $O(1)$ time.

Let $S$ be the string of two types of parentheses derived from $S_1$
and $S_2$ as follows. Let $\leftp{}$ and $\rightp{}$ be of type 1 with
$\leftb{}$ and $\rightb{}$ of type 2.  Initially, for each
$i=1,2,\ldots,2n$, replace the $i$-th $1$ of $S_2$ with
$S_1[i]$. Then, replace each $0$ of $S_2$ with a bracket such that the
bracket is open if and only if the last parenthesis in $S$ preceding
this $0$ is closed. More precisely, for each $i=1,2,\ldots,|S_2|$, let
\begin{displaymath}
  S[i]=
  \left\{
  \begin{array}{ll}
     S_1[j_1] &\text{if $S_2[i]=1$};\\
     \rightb{}&\text{if $S_2[i]=0$ and $S_1[j_i]=\leftp{}$};\\
     \leftb{} &\text{if $S_2[i]=0$ and $S_1[j_i]=\rightp{}$},
  \end{array}
  \right.
\end{displaymath}
where $j_i=\rank{S_2,i,1}$. For example, if $H$ and $T$ are as given
in Figure~\ref{fig:vibi-real}(a), then $S$ is as in
Equation~(\ref{eq:S}). It is easily determined that there exists an
auxiliary string $\aux_3$ such that any $O(\log n)$ consecutive
symbols of $S$ is obtainable from $S_1+S_2+\aux_3$ in $O(1)$ time: Let
$\ell=\floor{\frac{1}{4}\log_2 n}$.  Clearly, the content of
$S[i,i+\ell-1]$ can be uniquely determined by the concatenation $S'$
of $S_2[i,i+\ell-1]$ and $S_1[j,j+\ell-1]$ with
$j=\rank{S_2,i,1}$. Clearly, $S'$ is obtainable from
$S_1+S_2+\aux_1(S_2)$ in $O(1)$ time.  Since $S'$ has $4^\ell$
distinct values, we can precompute in $O(n)$ time a table $M$ with
$|M|=o(n)$ such that the content of $S[i,i+\ell-1]$ is obtainable
from $S'$ and $M$ in $O(1)$ time. Hence, it suffices to let
$\aux_3=M+\aux_1(S_2)$.


For each $i\in\setof{1,2,\ldots,n}$, let $L_i$ be the interval
$[\ell_i+1,\select{S_2,\rank{S_2,\ell_i,1}+1,1}-1]$ and $R_i$ be the
interval $[h_i+1,\select{S_2,\rank{S_2,h_i,1}+1,1}-1]$, where
$\ell_i=\select{S,i,\leftp{}}$ and $h_i=\match{S,\ell_i}$.  Let
$(v_i,v_j)$ and $(v_{i'},v_{j'})$, with $i<j$ and $i'<j'$, be two
unrelated edges of $H$ with respect to $T$.  Since $T$ is an orderly
spanning tree of $H$, one can see that if $(v_{i'},v_{j'})$ is
enclosed by the cycle of $H$ determined by $T$ and $(v_i,v_j)$, then
$h_i<h_{i'}<\ell_{j'}<\ell_j$.  It follows that $v_i$ and $v_j$, with
$i<j$, are adjacent in $H-T$ if and only if there exists an index
$\ell\in R_i$ with $\match{S,\ell}\in L_j$. Therefore, one can
determine whether $(v_i,v_j)$ is an unrelated edge of $H$ with respect
to $T$, by checking whether $i''\in R_i$ and $j''\in L_j$ hold, where
$(i'',j'')=\enclosek{2}{S,\select{S,\rank{S_2,h_i,1}+1,\leftp{}},\ell_j}$.
Therefore, the adjacency query is derivable from
$S_2+S+\aux_1(S_2)+\aux_1(S)$ in $O(1)$ time.

It is not difficult to see that the neighbors of a degree-$d$ node
$v_i$ can be listed from $S+\aux_1(S)$ in $O(d)$ time: If $v_i$ is not
the root of $T$, then the parent of $v_i$ is $v_j$, where $j$ is
computable by
$(j_1,j_2)=\enclose{S,\select{S,i,\leftp{}},\match{S,\select{S,i,\leftp{}}}}$
and $j=\rank{S,j_1,\leftp{}}$.  If $v_i$ is not a leaf of $T$, then
$v_{i+1}$ is the first child of $v_i$ in $T$.  If $v_j$ is the $t$-th
child of $v_i$ in $T$, then the $(t+1)$-st child of $v_i$ in $T$ is
$v_k$, where $k=\rank{S,1+\match{S,\select{S,j,\leftp{}}},\leftp{}}$.
If $t\leq |B_2(v_i)|$, then the $t$-th neighbor of $v_i$ in $B_2(v)$
with respect to $T$ is $v_j$, where $j$ is computable by
$j_1=\match{S,t+\select{S,i,\leftp{}}}$,
$j_2=\select{S,\rank{S,j_1,\rightp{}},\rightp{}}$,
$j=\rank{S,\match{S,j_2},\leftp{}}$.  If $t\leq |B_4(v_i)|$, then the
$t$-th neighbor of $v_i$ in $B_4(v)$ with respect to $T$ is $v_j$
where $j$ is computable by $j_1=\match{S,\select{S,i,\leftp{}}}$ and
$j=\rank{S,\match{S,j_1+t},\leftp{}}$.

It is not difficult to verify that $G$ can be reconstructed from $S$
and $S_3$ in $O(m+n)$ time.  Therefore, the theorem is proved by
letting $\aux = \aux'+\aux_3+\aux_1(S)$.
\end{proof}



\section*{Acknowledgments}
We thank Xin He for helpful comments. We thank Hsu-Chun Yen and Ho-Lin
Chen for presenting~\cite{FoessmeierKK97} to our attention.  We also
thank Richie Chih-Nan Chuang, Yuan-Jiunn Wang, and Kai-Ju Liu for
discussions.  
\bibliographystyle{abbrv} 
\bibliography{tree}

\begin{thebibliography}{10}

\bibitem{ArikatiMZ97}
S.~R. Arikati, A.~Maheshwari, and C.~D. Zaroliagis.
\newblock Efficient computation of implicit representations of sparse graphs.
\newblock {\em Discrete Applied Mathematics}, 78:1--16, 1997.

\bibitem{BCW90}
T.~C. Bell, J.~G. Cleary, and I.~H. Witten.
\newblock {\em Text Compression}.
\newblock Prentice-Hall, Englewood Cliffs, NJ, 1990.

\bibitem{BoseHS96}
P.~Bose, A.~M. Dean, and J.~P. Hutchinson.
\newblock On rectangle visibility graphs.
\newblock In North \cite{GD96}, pages 25--44.

\bibitem{Breuer66}
M.~Breuer.
\newblock Coding vertices of a graph.
\newblock {\em {IEEE} Transactions on Information Theory}, 12:148--153, 1966.

\bibitem{BreuerF67}
M.~Breuer and J.~Folkman.
\newblock An unexpected result on coding vertices.
\newblock {\em Journal of Mathematical Analysis and Applications}, 20:583--600,
  1967.

\bibitem{Brodal99e}
G.~S. Brodal and R.~Fagerberg.
\newblock Dynamic representations of sparse graphs.
\newblock In {\em Proceedings of the 6th International Workshop on Algorithms
  and Data Structures}, Lecture Notes in Computer Science 1663, pages 342--351.
  Springer-Verlag, 1999.

\bibitem{BrodnikM00}
A.~Brodnik and J.~I. Munro.
\newblock Membership in constant time and almost-minimum space.
\newblock {\em {SIAM} Journal on Computing}, 28(5):1627--1640, 2000.

\bibitem{ChenLLY02}
H.-L. Chen, C.-C. Liao, H.-I. Lu, and H.-C. Yen.
\newblock Some applications of orderly spanning trees in graph drawing.
\newblock In {\em Proceedings of the 10th International Symposium on Graph
  Drawing}, Lecture Notes in Computer Science, Irvine, California, August
  26--28 2002, to appear. Springer-Verlag.

\bibitem{ChrobakN98}
M.~Chrobak and S.-i. Nakano.
\newblock Minimum-width grid drawings of plane graphs.
\newblock {\em Computational Geometry: Theory and Applications}, 11(1):29--54,
  1998.

\bibitem{Chrobak95}
M.~Chrobak and T.~H. Payne.
\newblock A linear-time algorithm for drawing a planar graph on a grid.
\newblock {\em Information Processing Letters}, 54(4):241--246, May 1995.

\bibitem{ChuangGHKL98}
R.~C.-N. Chuang, A.~Garg, X.~He, M.-Y. Kao, and H.-I. Lu.
\newblock Compact encodings of planar graphs via canonical ordering and
  multiple parentheses.
\newblock In K.~G. Larsen, S.~Skyum, and G.~Winskel, editors, {\em Proceedings
  of the 25th International Colloquium on Automata, Languages, and
  Programming}, Lecture Notes in Computer Science 1443, pages 118--129,
  Aalborg, Denmark, 1998. Springer-Verlag.

\bibitem{Clark96}
D.~R. Clark.
\newblock {\em Compact {PAT} Trees}.
\newblock PhD thesis, University of Waterloo, 1996.

\bibitem{CohenDKT93}
R.~F. Cohen, G.~Di~Battista, A.~Kanevsky, and R.~Tamassia.
\newblock Reinventing the wheel: An optimal data structure for connectivity
  queries.
\newblock In {\em Proceedings of the 25th Annual ACM Symposium on the Theory of
  Computing}, pages 194--200, San Diego, California, 16--18~May 1993.

\bibitem{DeFPP90}
H.~{de~Fraysseix}, J.~Pach, and R.~Pollack.
\newblock How to draw a planar graph on a grid.
\newblock {\em Combinatorica}, 10:41--51, 1990.

\bibitem{DeanH97}
A.~M. Dean and J.~P. Hutchinson.
\newblock Rectangle-visibility representations of bipartite graphs.
\newblock {\em Discrete Applied Mathematics}, 75:9--25, 1997.

\bibitem{DeanH98}
A.~M. Dean and J.~P. Hutchinson.
\newblock Rectangle-visibility layouts of unions and products of trees.
\newblock {\em Journal of Graph Algorithms and Applications}, 2(8):1--21, 1998.

\bibitem{Dushnik41}
B.~Dushnik and E.~W. Miller.
\newblock Partially ordered sets.
\newblock {\em American Journal of Mathematics}, 63:600--610, 1941.

\bibitem{Elias75}
P.~Elias.
\newblock Universal codeword sets and representations of the integers.
\newblock {\em {IEEE} Transactions on Information Theory}, IT-21:194--203,
  1975.

\bibitem{FoessmeierKK97}
U.~F\"{o}\ss{}meier, G.~Kant, and M.~Kaufmann.
\newblock 2-visibility drawings of planar graphs.
\newblock In North \cite{GD96}, pages 155--168.

\bibitem{FredmanW94}
M.~L. Fredman and D.~E. Willard.
\newblock Trans-dichotomous algorithms for minimum spanning trees and shortest
  paths.
\newblock {\em Jouranl of Computer and System Sciences}, 48(3):533--551, June
  1994.

\bibitem{GW83}
H.~Galperin and A.~Wigderson.
\newblock Succinct representations of graphs.
\newblock {\em Information and Control}, 56:183--198, 1983.

\bibitem{GavoilleICALP99}
C.~Gavoille and N.~Hanusse.
\newblock Compact routing tables for graphs of bounded genus.
\newblock In J.~Wiedermann, P.~van Emde~Boas, and M.~Nielsen, editors, {\em
  Proceedings of the 26th International Colloquium on Automata, Languages, and
  Programming}, Lecture Notes in Computer Science 1644, pages 351--360, Prague,
  Czech Republic, 1999. Springer-Verlag.

\bibitem{GrossiL98}
R.~Grossi and E.~Lodi.
\newblock Simple planar graph partition into three forests.
\newblock {\em Discrete Applied Mathematics}, 84:121--132, 1998.

\bibitem{HarelS98}
D.~Harel and M.~Sardas.
\newblock An algorithm for straight-line drawing of planar graphs.
\newblock {\em Algorithmica}, 20(2):119--135, 1998.

\bibitem{He99}
X.~He.
\newblock On floor-plan of plane graphs.
\newblock {\em {SIAM} Journal on Computing}, 28(6):2150--2167, 1999.

\bibitem{HKL99}
X.~He, M.-Y. Kao, and H.-I. Lu.
\newblock A fast general methodology for information-theoretically optimal
  encodings for graphs.
\newblock In J.~Ne\v{s}et\v{r}il, editor, {\em Proceedings of the 7th Annual
  European Symposium on Algorithms}, Lecture Notes in Computer Science 1643,
  pages 540--549, Prague, Czech Republic, 16--18 July 1999. Springer-Verlag.

\bibitem{HKL}
X.~He, M.-Y. Kao, and H.-I. Lu.
\newblock Linear-time succinct encodings of planar graphs via canonical
  orderings.
\newblock {\em {SIAM} Journal on Discrete Mathematics}, 12(3):317--325, 1999.

\bibitem{HKL2000}
X.~He, M.-Y. Kao, and H.-I. Lu.
\newblock A fast general methodology for information-theoretically optimal
  encodings of graphs.
\newblock {\em {SIAM} Journal on Computing}, 30(3):838--846, 2000.

\bibitem{Hoprcoft74}
J.~Hoprcoft and R.~E. Tarjan.
\newblock Efficient planrity testing.
\newblock {\em Journal of the ACM}, 21(4):549--568, 1974.

\bibitem{HutchinsonSV99}
J.~P. Hutchinson, T.~Shermer, and A.~Vince.
\newblock On representations of some thickness-two graphs.
\newblock {\em Computational Geometry: Theory and Applications},
  13(3):161--171, 1999.

\bibitem{FOCS38}
IEEE.
\newblock {\em Proceedings of the 38th Annual Symposium on Foundations of
  Computer Science}, Miami Beach, Florida, 20--22 Oct. 1997.

\bibitem{IR82}
A.~Itai and M.~Rodeh.
\newblock Representation of graphs.
\newblock {\em Acta Informatica}, 17:215--219, 1982.

\bibitem{Jacobson89}
G.~Jacobson.
\newblock Space-efficient static trees and graphs.
\newblock In {\em Proceedings of the 30th Annual Symposium on Foundations of
  Computer Science}, pages 549--554, Research Triangle Park, North Carolina, 30
  Oct.--1 Nov. 1989. IEEE.

\bibitem{KNR92}
S.~Kannan, M.~Naor, and S.~Rudich.
\newblock Implicit representation of graphs.
\newblock {\em {SIAM} Journal on Discrete Mathematics}, 5:596--603, 1992.

\bibitem{Kant96}
G.~Kant.
\newblock Drawing planar graphs using the canonical ordering.
\newblock {\em Algorithmica}, 16(1):4--32, 1996.

\bibitem{KantH98}
G.~Kant and X.~He.
\newblock Regular edge labeling of $4$-connected plane graphs and its
  applications in graph drawing problems.
\newblock {\em Theoretical Computer Science}, 172(1-2):175--193, 1997.

\bibitem{kaot94.isaac}
M.-Y. Kao and S.~H. Teng.
\newblock Simple and efficient compression schemes for dense and complement
  graphs.
\newblock In {\em Proceedings of the 5th Annual Symposium on Algorithms and
  Computation}, Lecture Notes in Computer Science 834, pages 201--210, Beijing,
  China, 1994. Springer-Verlag.

\bibitem{KW:encodings}
K.~Keeler and J.~Westbrook.
\newblock Short encodings of planar graphs and maps.
\newblock {\em Discrete Applied Mathematics}, 58:239--252, 1995.

\bibitem{LiaoLY01}
C.-C. Liao, H.-I. Lu, and H.-C. Yen.
\newblock Floor-planning via orderly spanning trees.
\newblock In {\em Proceedings of the 9th International Symposium on Graph
  Drawing}, Lecture Notes in Computer Science 2265, pages 367--377, Vienna,
  Austria, September 23--26 2001. Springer-Verlag.

\bibitem{Lu02:SODA}
H.-I. Lu.
\newblock Linear-time compression of bound-genus graphs into
  information-theoretically optimal number of bits.
\newblock In {\em Proceedings of the 13th Annual {ACM}-{SIAM} Symposium on
  Discrete Algorithms}, pages 223--224, San Francisco, 6--8 Jan. 2002. ACM and
  SIAM.

\bibitem{Lu02}
H.-I. Lu.
\newblock Improved compact routing tables for planar networks via orderly sp
  anning trees.
\newblock In {\em Proceedings of the 8th Annual International Computing and
  Combinatorics Conference}, Lecture Notes in Computer Science 2387, Singapore,
  August 15--17 2002, to appear. Springer-Verlag.

\bibitem{MR97}
J.~I. Munro and V.~Raman.
\newblock Succinct representation of balanced parentheses, static trees and
  planar graphs.
\newblock In {\em Proceedings of the 38th Annual Symposium on Foundations of
  Computer Science\/} \cite{FOCS38}, pages 118--126.

\bibitem{MunroRS01}
J.~I. Munro, V.~Raman, and A.~Storm.
\newblock Representing dynamic binary trees succinctly.
\newblock In {\em Proceedings of the 12th Annual {ACM}-{SIAM} Symposium on
  Discrete Algorithms}, pages 529--536, Washington, DC, 7--9 Jan. 2001.

\bibitem{naor90}
M.~Naor.
\newblock Succinct representation of general unlabeled graphs.
\newblock {\em Discrete Applied Mathematics}, 28:303--307, 1990.

\bibitem{GD96}
S.~North, editor.
\newblock {\em Proceedings of the 4th International Symposium on Graph
  Drawing}, Lecture Notes in Computer Science 1190, California, USA, 1996.
  Springer-Verlag.

\bibitem{PH86.encode}
C.~H. Papadimitriou and M.~Yannakakis.
\newblock A note on succinct representations of graphs.
\newblock {\em Information and Control}, 71:181--185, 1986.

\bibitem{Rossignac99}
J.~Rossignac.
\newblock Edgebreaker: Connectivity compression for triangle meshes.
\newblock {\em {IEEE} Transactions on Visualization and Computer Graphics},
  5(1):47--61, 1999.

\bibitem{Schnyder89}
W.~Schnyder.
\newblock Planar graphs and poset dimension.
\newblock {\em Order}, 5:323--343, 1989.

\bibitem{Schnyder90}
W.~Schnyder.
\newblock Embedding planar graphs on the grid.
\newblock In {\em Proceedings of the First Annual {ACM}-{SIAM} Symposium on
  Discrete Algorithms}, pages 138--148, 1990.

\bibitem{Spinrad-book}
J.~Spinrad.
\newblock {\em Efficient Representation of Graphs}.
\newblock http://www.vuse.vanderbilt.edu/ \~{ }spin/research.html, in
  preparation.

\bibitem{Talamo-Vocca-Compact-1998}
M.~Talamo and P.~Vocca.
\newblock Compact implicit representation of graphs.
\newblock In J.~Hromkovic and O.~S\'{y}kora, editors, {\em Proceedings of the
  Graph-Theoretic Concepts in Computer Science}, Lecture Notes in Computer
  Science 1517, pages 164--176. Springer, Smolenice Castle, Slovak Republic,
  1998.

\bibitem{Thorup97}
M.~Thorup.
\newblock Undirected single source shortest paths in linear time.
\newblock In {\em Proceedings of the 38th Annual Symposium on Foundations of
  Computer Science\/} \cite{FOCS38}, pages 12--21.

\bibitem{Thorup2000}
M.~Thorup.
\newblock On {RAM} priority queues.
\newblock {\em {SIAM} Journal on Computing}, 30:86--109, 2000.

\bibitem{trotter92}
W.~T. Trotter.
\newblock {\em Combinatorics and Partially Ordered Sets --- Dimension Theory}.
\newblock Johns Hopkins University Press, Baltimore, MD, 1992.

\bibitem{turan84}
G.~Tur\'{a}n.
\newblock On the succinct representation of graphs.
\newblock {\em Discrete Applied Mathematics}, 8:289--294, 1984.

\bibitem{Tutte62}
W.~T. Tutte.
\newblock A census of planar triangulations.
\newblock {\em Canadian Journal of Mathematics}, 14:21--38, 1962.

\bibitem{Tutte63b}
W.~T. Tutte.
\newblock A census of planar maps.
\newblock {\em Canadian Journal of Mathematics}, 15:249--271, 1963.

\bibitem{Handbook}
P.~{van Emde Boas}.
\newblock Machine models and simulations.
\newblock In J.~{van Leeuwen}, editor, {\em Handbook of Theoretical Computer
  Science}, volume~A, chapter~1, pages 1--60. Elsevier, Amsterdam, 1990.

\bibitem{Yannakakis89}
M.~Yannakakis.
\newblock Embedding planar graphs in four pages.
\newblock {\em Jouranl of Computer and System Sciences}, 38(1):36--67, Feb.
  1989.

\bibitem{YeapS93}
K.-H. Yeap and M.~Sarrafzadeh.
\newblock Floor-planning by graph dualization: $2$-concave rectilinear modules.
\newblock {\em SIAM Journal on Computing}, 22(3):500--526, 1993.

\end{thebibliography}
\end{document}